\documentclass[conference]{IEEEtran}

\newcommand{\sharma}[2]{\noindent\textcolor{blue}{~#1~}\textcolor{red}{\textbf{(Sharma):}~#2 \\}} %for displaying Sharma's comments

\usepackage{comment}
\usepackage{url}
\usepackage{algorithm}
\usepackage{algorithmic}
\usepackage{nth}
\usepackage{amssymb}
\usepackage{epsfig}
\usepackage{wrapfig}
\usepackage{enumitem}
\usepackage{subfig}
\usepackage{color}
\usepackage{multirow}
\usepackage{graphicx}

\usepackage[mathscr]{euscript}

\ifCLASSOPTIONcompsoc
  \usepackage[nocompress]{cite}
\else
  \usepackage{cite}
\fi
\ifCLASSINFOpdf
\else
\fi
\usepackage{amsmath}

\begin{document}

% to add page numbers - REMOVE BEFORE SUBMISSION

%\thispagestyle{plain}
%\pagestyle{plain}

\title{Structure-Preserving Community In A Multilayer Network: Definition, Detection, And Analysis}
\title{Efficient Approach to Compute Structure- And Semantics-Preserving Community In a Heterogeneous Multilayer Network}
\title{Computing Structure- And Semantics-Preserving Community in a Heterogeneous Multilayer Network}
\title{An Efficient Framework for Computing Structure- And Semantics-Preserving Community in a Heterogeneous Multilayer Network}

%\author{Abhishek~Santra,        Sanjukta~Bhowmick        and~Sharma~Chakravarthy}

\author{\IEEEauthorblockN{Abhishek Santra\IEEEauthorrefmark{1},
Kanthi Sannappa Komar\IEEEauthorrefmark{2}, Sanjukta Bhowmick\IEEEauthorrefmark{3} and
Sharma Chakravarthy\IEEEauthorrefmark{4}}
\IEEEauthorblockA{\IEEEauthorrefmark{1}\IEEEauthorrefmark{2}\IEEEauthorrefmark{4}IT Lab and CSE Department, University of Texas at Arlington, Arlington, Texas \\
\IEEEauthorrefmark{3}CSE Department, University of North Texas, Denton, Texas \\
Email: \IEEEauthorrefmark{1}abhishek.santra@mavs.uta.edu,
\IEEEauthorrefmark{2}kanthisannappa.komar@mavs.uta.edu,\\
\IEEEauthorrefmark{3}sanjukta.bhowmick@unt.edu,
\IEEEauthorrefmark{4}sharma@cse.uta.edu}}

\IEEEtitleabstractindextext{%

\begin{abstract}
%%Modeling and analysis of complex data sets (i.e., data sets with multiple entity and feature  types)  are challenging especially if one wants to do it in a \textit{flexible and efficient} manner to match the analysis requirements. Analysis entails understanding of the data set with respect to entity and feature semantics as well as inferring useful knowledge through analysis. We posit that modeling of these data sets can be done elegantly using multilayer networks or MLNs (also called multiplexes, layers of networks that are inter-connected) instead of using a single graph (or attribute graph.) From a big data analysis perspective, efficient and scalable approaches are needed for this representation. 

%%In this paper,  we first illustrate the elegance of multilayer networks for modeling by using two well-known data sets, namely, the IMDb  and the DBLP. Modeling alternatives will be analyzed leading to the need for heterogeneous multiplexes, the main focus of this paper. 

Multilayer networks or MLNs (also called  multiplexes or network of networks) are being used extensively for modeling and analysis of data sets with multiple entity and feature  types and associated relationships. Although the concept of community is widely-used for aggregate analysis, a \textbf{structure- and semantics-preserving} definition for it is lacking for MLNs. Retention of original \textit{MLN structure} and \textit{entity relationships} is important for detailed drill-down analysis. In addition, efficient computation is also critical for large number of analysis.
%%There is no structure-preserving definition of a community for a MLN as most of the current analyses aggregate a MLN to a single graph.
%%\textcolor{blue}{SB:There is no consensus for community definition actually. What is structure preserving, why do we need it. I think we need to say that in the beginning}

%%Although there is consensus on community definition for single graphs (and detection packages) and to a lesser extent for homogeneous MLNs, it is lacking for heterogeneous MLNs. 
In this paper, we introduce a structure-preserving community definition for MLNs as well as a framework for its efficient computation using the decoupling approach.
The proposed decoupling approach combines communities from individual layers to form a \textit{serial k-community} for connected k layers in a MLN. We propose a new algorithm for pairing communities across layers and introduce several weight metrics for composing communities from two layers using participating community characteristics. In addition to the definition, our proposed approach has a number of desired characteristics. It: i) leverages extant single graph community detection algorithms, ii) introduces several weight metrics that are customized for the community concept,  iii) is a new algorithm  for pairing communities using bipartite graphs,  and iv) experimentally validates the community computation and its efficiency on widely-used IMDb  and DBLP data sets.

%%\textcolor{blue}{SB:Not all of these are "advantages", some are features and some are contributions.}
%%, we validate our proposed decoupling-based approach for analysis, efficiency, and flexibility for user-driven analysis.\\~\\
%more than of allow the flexible analysis of such data sets by the mix-and-match of different layers. Moreover, we have defined the different categories of community-based analysis on whose basis we have proposed the flow and composition based approaches for community detection using bipartite graphs. The applicability of our approach has been demonstrated on a movie dataset (IMDb data set) composed of networks of actors, movies and directors.

\end{abstract}

\begin{IEEEkeywords}
Heterogeneous Multilayer Networks; Bipartite Graphs; Community Definition and Detection; Decoupling-Based Composition
\end{IEEEkeywords}}

\maketitle

% is not selected <OR> if conference mode is selected - because compsoc
% conference papers position the abstract like regular (non-compsoc)
\IEEEdisplaynontitleabstractindextext

% \fi
\IEEEpeerreviewmaketitle



\section{Motivation}
\label{sec:introduction}
As data sets become more complex in terms of entity and feature types, the approaches needed for their modeling and analysis also warrant extensions or new alternatives to match the data set complexity. With the advent of social networks and large data sets, we have already seen a surge in the use of graph-based modeling along with a renewed interest in concepts, such as community and hubs used for their analysis. 

Informally, MLNs\footnote{The terminology used for variants of multilayer networks varies drastically in the literature and many a times is not even consistent with one another. For clarification, please refer to~\cite{MultiLayerSurveyKivelaABGMP13} which provides an excellent comparison of terminology used in the literature, their differences, and usages clearly.} are \textit{layers of networks} where each layer is a simple graph and captures the semantics of an attribute (or feature) of an entity type using an edge to represent that relationship. The layers can also be connected. If each layer of a MLN has the \texttt{same set of entities of the same type}, it is termed a homogeneous MLN (or HoMLN.) For a HoMLN, intra-layer edges are shown explicitly and inter-layer edges are implicit (and not shown.) If \texttt{the set and types of entities are different for each layer}, then relationships of entities across layers are also shown using explicit inter-layer edges. This distinguishes a heterogeneous MLN (or HeMLN) from the previous one. 
%%Of course, hybrid MLNs (or HyMLN) are also possible.

%%%The contribution of this paper is a new definition  for a k-community in a HeMLN. Unlike others, it is structure-preserving\footnote{By this we imply that the communities detected using our definition preserve their multilayer network structure (including inter- and intr-layer edges) and node/edge labels and types. In other words, each community  is a heterogeneous MLN in its own right!},  efficient from a computation perspective, and expressive from an analysis perspective. Our use of bipartite graphs for defining communities across layers is different from the approaches found in the literature, such as tensors~\cite{LayerAggDomenicoNAL14,de2013mathematical}, adjacency matrices~\cite{DeDomenico201318469,articleGomez}, and others. In addition, we  formalize a decoupling approach that: i) uses extant single graph algorithms for each layer, ii) composes partial results from individual layers to obtain larger combined communities (termed \textit{a serial k-community}), iii) customizes the semantics of communities using metrics, and iv) preserves MLN structure of the communities unlike extant approaches. For this we use a bipartite graph approach for composition, Furthermore, this approach is efficient due to decoupling  and is scalable as it is amenable to parallelization. Finally, this approach supports flexible analysis of any subset of features (or layers) avoiding re-computation of communities at the layer level.

\subsection{Structure- and Semantics-Preservation}
\label{sec:structure-preservation}
For a simple graph, a community preserves its structure in terms of node/edge labels and relationships. Preserving the structure of a community of a MLN (especially HeMLN) entails preserving their multilayer network structure and preserving semantics includes preserving node/edge types, labels, and importantly inter-layer relationships. In other words, each community should be a heterogeneous MLN in its own right!
Current approaches, such as using the MLN as a whole~\cite{Wilson:2017:CEM:3122009.3208030}, type-independent~\cite{LayerAggDomenicoNAL14}, and projection-based~\cite{Berenstein2016, sun2013mining}, do not accomplish this as they aggregate (or collapse) layers into a simple graph in different ways. Importantly, aggregation approaches are likely to result in some information loss~\cite{MultiLayerSurveyKivelaABGMP13}, distortion of properties~\cite{MultiLayerSurveyKivelaABGMP13}, or hide the effect of different entity types and/or different intra- or inter-layer relationships as elaborated in~\cite{DeDomenico201318469}. Structure-preservation is critical for understanding a HeMLN community and for drill-down or detailed analysis of communities.

Without structure- and semantics-preservation, it is not possible to understand the result of the analysis as mapping to original node labels and relationships is extremely difficult (or even not possible) when more than several layers are involved. We will demonstrate from our experimental results how easily we can drill down to see patterns in terms of original labels.
%%(e.g., \textit{actor and director names for the IMDb data set and authors groups widely publishing in specific conferences for the DBLP data set}.)

\subsection{Decoupling Approach For Efficiency}
Decoupling as proposed in this paper is the equivalent of ``divide and conquer" for MLNs. Research on modeling a data set as a MLN \textit{and} computing on the whole MLN has not addressed efficiency issues. As with divide and conquer, decoupling requires partitioning (which comes from the MLN structure) and a way to compose partial (or intermediate) results. This paper uses a customized bipartite graph match as the composition function (referred to as $\Theta$, in this paper) leading to efficient community detection on MLNs. 
%%Decoupling approach has also been shown to be beneficial for community detection on HoMLNs~\cite{ICCS/SantraBC17} and other computations, such as hubs, in ~\cite{ICDMW/SantraBC17}.
%Our approach for a multiplex is to compose communities of individual layers to preserve the semantics of the community for the combination of layers. This has been addressed in the literature for \textit{homogeneous} multiplexes. Hence, this paper focuses on community detection for \textit{heterogeneous} multiplexes using communities from individual layers. 
The contributions of this paper are:
 
\begin{itemize}
%\item Analyzing alternative modeling approaches for complex data sets from four domains with multiple entity and features types,
%\item Establishing the advantages of MLNs for modeling in general and HeMLNs in particular for further analysis,
\item Definition of structure- and semantics-preserving k-community for a HeMLN (Section~\ref{sec:hemln-community}),
\item Identification of a composition function and formalizing decoupling-based approach for k-community detection with an algorithm  (Section~\ref{sec:community-detection}),
\item Mapping of detailed analysis requirements of the data set using the k-community and weight choices (Section~\ref{sec:application-and-analysis}),
\item A new bipartite match algorithm (termed Maximum Weighted Bipartite Coupling or MWBC) for composing layers and identification of useful weight metrics and their uniqueness (Section ~\ref{sec:customize-maxflow}), and
\item Experimental analysis using the IMDb and DBLP data sets to establish the validity of the proposed approach along with performance analysis  (Section~\ref{sec:experiments}.)
\end{itemize}

The paper is organized as indicated above with related work in Section~\ref{sec:related-work} and conclusions in Section~\ref{sec:conclusions}.

\section{Related Work}
\label{sec:related-work}

As the focus of this paper is community definition and its efficient detection in HeMLNs, we present relevant work on simple graphs and MLNs. The advantages of modeling using MLNs are discussed in~\cite{Boccaletti20141, MultiLayerSurveyKivelaABGMP13}.
%% BDA/SantraB17,,CommSurveyKimL15

\textit{Community detection} on a simple graph involves identifying groups of vertices that are more connected to each other than to other vertices in the network. Most of the  work in the literature considers \textbf{single networks or simple graphs} where this objective is translated to optimizing network parameters such as modularity ~\cite{clauset2004} or conductance ~\cite{Leskovec08}. As the combinatorial optimization of community detection is NP-complete ~\cite{Brandes03}, a large number of competitive approximation algorithms have been developed (see reviews in ~\cite{ Xie2013}.) %%Fortunato2009,
Algorithms for community detection have been developed for different types of input graphs including directed ~\cite{Yang10}, %%, Leicht08
edge-weighted ~\cite{Berry2011}, and dynamic networks~\cite{porterchaos13}. %%,Bansal201
%Recently there have also been algorithms for identifying overlapping communities~\cite{Yang2013acm,Chakraborty2015a}.
However, to the best of our knowledge, there is no  community definition and detection that include node and edge labels, node weights as well as graphs with self-loops and multiple edges between nodes\footnote{This is in contrast to subgraph mining, search, and querying of graphs where non-simple or attributed graphs are widely used.}.
%%In contrast, subgraph mining~\cite{datamine/KuramochiK05,KDD/HolderCD1994, tkde/DasC18}, querying~\cite{tkde/JayaramKLYE15,dawak/DasGC16}, and search~\cite{bigdataconf/HaoC0HBH15,pods/ShashaWG02} have used graphs with node and/or edge labels including multiple edges between nodes, cycles, and self-loops.}. 
Even the most popular community detection packages such as Infomap \cite{InfoMap2014} or Louvain~\cite{DBLP:Louvain}, do not accept non-simple graphs. 

Recently, community detection algorithms have been extended to \textbf{HoMLNs} (see reviews \cite{CommSurveyKimL15,CommFortunatoC09}.)  Algorithms based on matrix factorization, % \cite{dong2012clustering}
cluster expansion philosophy,  %\cite{li2008scalable}
Bayesian probabilistic models,  %\cite{xu2012model}
regression, %\cite{cai2005mining}
and spectral optimization of the modularity function based on the supra-adjacency representation %\cite{zhang2017modularity}
%and a significance based score that quantifies the connectivity of an observed vertex-layer set through comparison with a fixed degree random graph model% \cite{wilson2017community}
have been developed.
% chnaged above para
However, all these approaches \textit{analyze a MLN either by aggregating all (or a subset of) layers of a HoMLN using Boolean and other operators or by considering the entire MLN as a whole}. 
%%Recently a decoupling-based approaches for detecting communities~\cite{ICCS/SantraBC17} and centrality~\cite{ICDMW/SantraBC17} in HoMLN have been proposed. 
%The decoupling approach uses partial results from individual layers and  composes them for Boolean operators to compute communities for combinations of layers. This approach is more efficient as it avoids re-computation of layer communities and provides  flexibility for analysis. This approach nicely leverages the algorithms developed for a single network. 
%%They reduce the exhaustive analysis complexity from O($2^N$) (for all possible subsets of layers) to linear complexity for \textit{N} layers by composing combinations of them. The goal of this paper is to do the same for HeMLNs.

%allows the flexible analysis of any complex dataset modeled as a homogeneous multilayer network in a cost-effective manner as one does need to to generate and separately analyze the combined networks (close to O($2^N$) in an exhaustive analysis, if the multiplex has N layers).

Majority of the work on analyzing HeMLN (reviewed in \cite{shi2017survey,sun2013mining}) focuses on developing meta-path based techniques for determining the similarity of objects~\cite{wang2016relsim}, classification of objects~\cite{wang2016text}, predicting the missing links~\cite{zhang2015organizational}, ranking/co-ranking~\cite{shi2016constrained} and recommendations~\cite{shi2015semantic}. An important aspect to be noted here is that most of them do not consider the intra-layer relationships and concentrate mainly on the bipartite graph formed by the inter-layer edges. %Further, they either consider the multilayer network as a whole or remove the type information of nodes or project one layer onto another while proposing the approaches, thus neglecting the effect of different combinations

%%In contrast to the homogeneous case, there \textbf{is no consensus on the definition of a community, let alone structure-preserving} for \textbf{HeMLNs}. 
The type-independent~\cite{LayerAggDomenicoNAL14} and projection-based~\cite{Berenstein2016} approaches used for HeMLNs \textit{do not preserve the structure or semantics of the community.} The type independent approach collapses  all layers into a simple graph keeping \textit{all} nodes and edges (including inter-layer edges) sans their types and labels. The same is true for the projection-based approach as well.
%%projects the nodes of one layer onto another layer and uses the layer neighbor and inter-layer edges to collapse the two layers into a single graph with a single entity type instead of two. 
The presence of different sets of entities in each layer and the presence of intra-layer edges makes structure-preserving definition more challenging for HeMLNs and also warrants a novel composition technique (as proposed in this paper.) A few existing works have proposed techniques for generating clusters of entities \cite{melamed2014community}, but they have only considered the inter-layer links and not the networks themselves.
This paper hopes to fill the gap for a \textit{structure- and semantics-preserving community}.
%%Thus, the \textit{combined effect of layer communities, entity types, intra- and inter-layer relationships (types) have not been included in defining a community in a HeMLN}.  

%%This is very much needed for mapping analysis requirements that require community-based computations when a data set is modeled as a heterogeneous MLN.

%This paper proposes a decoupling approach to community detection in HeMLNs. As in~\cite{ICCS/SantraBC17}, the goal is to use partial results from each layer and compose them to detect communities for a given set of layers. As we demonstrate, unlike the homogeneous MLNs, the analysis semantics plays a role in the composition of layers in terms of the choice of metrics used for composition. 
%In this paper, we will \textit{focus on proposing decoupling-based approaches for determining heterogeneous communities}.
\section{Definitions}
\label{sec:problem-statement}

A {\bf graph} $G$ is an ordered pair $(V, E)$, where $V$ is a set of vertices and $E$ is a set of edges. An edge $(u,v)$ is a 2-element subset of the set $V$. The two vertices that form an edge are said to be adjacent or neighbors. In this paper we only consider graphs that are undirected.
%%(the  vertices in the edge are unordered) and simple (there are no self-loops or multiple edges.) Although this type of graph does not include labels, we note that our representation and analysis can be applied to labeled graphs.

%We also do not consider any labels on the vertices or edges, except for weights (real numbers) associated with edges when conducting the bi-partite matching.

%%\textcolor{blue}{SB:We have not talked about layers yet--so may be define layers later. I dont think we have formally defined MLN}

%\abhi{2/14/2014}{Discuss this para. single networks?}

A {\bf multilayer network}, $MLN (G, X)$, is defined by two sets of graphs: i) The set $G = \{G_1, G_2, \ldots, G_N\}$ contains graphs of N individual layers as defined above, where $G_i (V_i, E_i)$ is defined by a set of vertices, $V_i$ and a set of edges, $E_i$. An edge $e(v,u) \in E_i$, connects vertices $v$ and $u$, where $v,u\in V_i$ and ii) A set $X =\{X_{1,2}, X_{1,3}, \ldots, X_{N-1,N}\}$ consists of bipartite graphs. Each graph $X_{i,j} (V_i, V_j, L_{i,j})$ is defined by two sets of vertices $V_i$ and $V_j$, and a set of edges (also called links or inter-layer edges) $L_{i,j}$, such that for every link $l(a,b) \in L_{i,j}$,  $a\in V_i$ and $b \in V_j$, where $V_i$ ($V_j$) is the vertex set of graph $G_i$ ($G_j$.)
%fix graph def

%%{\em To summarize}, the set $G$, represents the constituent graphs of layers of a MLN and the set $X$ represents the bipartite graphs that correspond to the edges \textit{between} pairs of layers (or inter-layer edges.)
%%The two sets of nodes for any bipartite graph from $X$ come from two layers and if labels are assigned, they would have different node labels.
For a HeMLN, $X$ is explicitly specified.
%given which is the focus of this paper.
Without loss of generality, we assume unique numbers for nodes across layers and disjoint sets of nodes across layers\footnote{Heterogeneous MLNs can also be defined with overlapping nodes across layers (see \cite{MultiLayerSurveyKivelaABGMP13}) which is not considered  in this paper.}.

%\abhi{2/22/2019}{This should be rephrased as k unique out of m layers (in a sequence) layers make a k-community}
%%\abhi{2/28/2019}{combining or composing ?}
We  propose a {\bf decoupling approach for HeMLN community detection}. Our algorithm is defined for combining communities from two layers of a HeMLN using a composition function and is extended to $k$ layers (by applying pair-wise composition repeatedly.) We define a \textit{serial k-community} to be a multilayer community where communities from $k$ distinct connected layers of a HeMLN are combined in a specified order.
%using the maximal matching approach to produce \textit{community pairs (or 2-communities)}. Further composition is performed on community pairs obtained for each layer pair using a join on 2-communities using community identity to produce the \textit{3-communities)}. This process is continued as needed to generate k-communities. Details of the approach are discussed in Section~\ref{sec:hemln-community}.

%sc: abhishek, need to slightly modify this fig. drop current approach. show pair-wise composition using bipat approach and join between them.
%SC: done

% metric based maximal flow matching
% meta node bipartite graph
% join based community detection
% no order
%$$\sharma{5/11/19}{Abhishek, check the expression. Also, add psi and theto to the diagram as we did in the asonam paper; HeMLN only}

\begin{figure}[h]
   \centering \includegraphics[width=\linewidth]{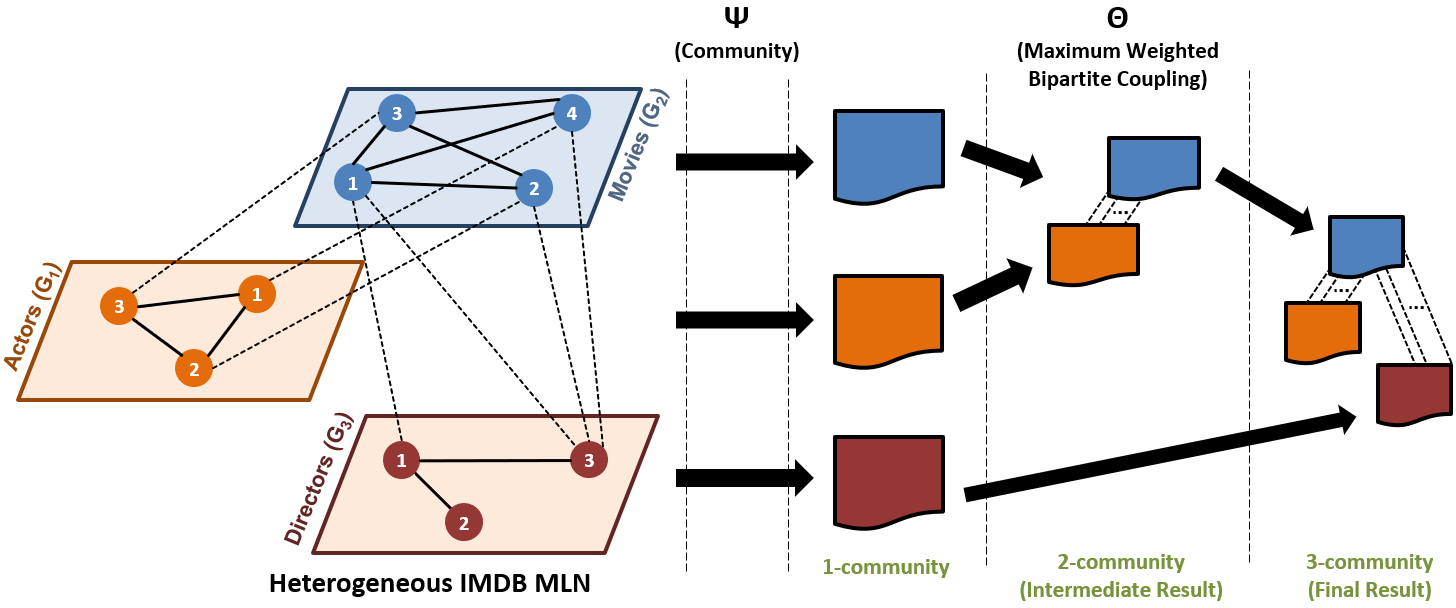}
   %\caption{Illustration of a serial 3-community definition for the sequence - (A $\Theta$ M)  $\Theta$ D \footnote{Technically, this should be expressed as. However, we use simpler notation for clarity.}}
   \caption{Illustration of decoupling approach for computing a 3-community (($G_2$ $\Theta_{2,1}$ $G_1$)  $\Theta_{2,3}~G_3$); $\omega_e$\protect\footnotemark}
   \label{fig:decoupling}
\vspace{-15pt}
\end{figure}

\footnotetext{Technically, this should be expressed as (($\Psi$($G_2$) $\Theta_{2,1}$ $\Psi$($G_1$))  $\Theta_{2,3}$ $\Psi$($G_3$).) However, we drop $\Psi$ for simplicity. In fact, $\Theta$ with its subscripts is sufficient for our purpose due to pre-defined precedence (left-to-right) of $\Theta$. We retain G for clarity of the expression. $\omega_e$ is a weight metric discussed in Section~\ref{sec:customize-maxflow}.}

%sc: this para repeats what is said in the prev para. need to fix.
%prev para should just explain the fig without going into details.details should be here.

%\abhi{2/14/2019}{discuss definition}

%%\textcolor{blue}{SB: I changed it finding communities, than general analysis, since that is the focus of the paper.}
Our proposed decoupling approach for finding HeMLN communities is as follows;

 {\bf(i)} First use the function $\Psi$ (here community detection) to find communities in each of the layers individually,% and consider each community to be a meta node,

 {\bf(ii)} for any two chosen layers, construct a bipartite graph using their communities as meta nodes and create meta edges that connect the meta nodes (using an appropriate element of X) and assign weight ($\omega$).  $\omega$ reflects the number of edges constituting a meta edge as well as properties of participating communities as discussed in Section~\ref{sec:customize-maxflow}, and

 {\bf(iii)} compose the partial results from each layer by representing each community as a meta node of the bipartite graph and using a function $\Theta$ which computes 2-community as pairs using the  \textit{weight information of edges in the bipartite graph}.)

 %%Although in this paper we focus on community detection, the functions $\Psi$, $\omega$ and $\Theta$ can be suitably modified for other network analysis, such as centrality detection (or hub), as well.

 Figure~\ref{fig:decoupling} illustrates the decoupling approach for specifying and computing a community of a larger size from partial results. It illustrates how a set of distinct communities from a layer is used for computing a 2-community (for 2 layers) and further a 3-community (for 3 layers) using partial results. 1-community is the set of communities generated for a layer (simple graph.)
We can define our problem statement as follows;
\textit{For a given data set with $\mathscr{F}$ different features and $\mathscr{T}$ entity types and a set of analysis objectives $\mathscr{A}$, model the data set using a HeMLN and determine the appropriate triad of $\Psi$,  $\Theta$, and $\omega$, and a $k$-community expression for computing \textit{each objective}.}

\section{Community Definition for a H\MakeLowercase{e}MLN}
\label{sec:hemln-community}

We first motivate the need for defining a  structure- and semantics-preserving communities.
For the IMDb data set, consider the HeMLM shown in Figure~\ref{fig:decoupling} and the analysis \textit{``Find groups of actors for every director group such that the most versatile members interact?} Note that the actor and director layers can only compute groups of actors and directors, who act in or direct similar genre, respectively. The connection (or coupling) between directors and actors only come from inter-layer edges. It is only by preserving the structure of both the communities in actor and director as well as the inter-layer edges, can we compute the answer that indicates the semantics of which actor groups are paired with the director groups. The inter-layer edges preserve the relationships of individual actors and directors as well.

\begin{table}[h]
    \renewcommand{\arraystretch}{1.3}
    %\vspace{-10pt}
    \centering
    % Some packages, such as MDW tools, offer better commands for making tables
    % than the plain LaTeX2e tabular which is used here.
        \begin{tabular}{|c|p{5.6cm}|}
            \hline
            $G_i (V_i, E_i)$ & Simple Graph for layer \textit{i} \\
            \hline
            $X_{i,j} (V_i,V_j,L_{i,j})$ & Bipartite graph of layers \textit{i} and \textit{j} \\%$G_i$ and $G_j$ with set $L_{i,j}$ of inter-layer edges\\
            \hline
            $MLN (G, X)$ & Multilayer Network of layer graphs (set \textit{G}) and Bipartite graphs (set \textit{X})\\
            \hline
            %%$F$ & Set of features (explicit or derived)\\             %%\hline
            %%$T$ & Set of entity types\\
            %%\hline
            $\Psi$ & Analysis function for $G_i$ (community)\\
            \hline
            $\Theta_{i,j}$ & Maximum Weighted Bipartite Coupling (MWBC) function \\
            %for partial results ($\Psi (G_i)$ and $\Psi (G_j)$)\\
            \hline
            $CBG_{i,j}$ & Community bipartite graph for $G_i$ and $G_j$ \\
            \hline
            $U_i$ & Meta nodes for layer \textit{i} 1-community \\
            \hline
            $L'_{i,j}$ & Meta edges between $U_i$ and $U_j$ \\
            \hline
            $c_i^{m}$ & $m^{th}$ community of $G_i$ \\
            \hline
            $v_i^{c^{m}}$, $e_i^{c^{m}}$ & Vertices and Edges in community $c_i^{m}$\\
            \hline
            $H_i^m$ & Hubs in $c_i^{m}$ \\
            \hline
            $H_{i,j}^{m,n}$ & Hubs in $c_i^{m}$ connected to $c_j^{n}$\\
            \hline
            $x_{i,j}$ & \{Expanded(meta edge $<c_i^{m}$, ~$c_j^{n}>$)\}\\
            \hline
            $0$ and $\phi$ & null community id and empty $x_{i,j}$\\
            \hline
            $\omega_e$, $\omega_d$, $\omega_h$ & Weight metrics for meta edges\\
            \hline
%            $L_{i,j}^{m,n}$ & Set of inter-layer edges b/w $c_i^{m}$ \& $c_j^{n}$ \\
 %           \hline

        \end{tabular}
            \caption{Notations used in this paper}
    \label{table:notations}
%\vspace{-20pt}
\end{table}

%%should identify pairs of communities from each layer as well as inter-layer edges that strongly couple the two sets of communities. Efficiency is critical if the community involves several layers that are large.

Clearly, multiple relationships can exist in such a collection of layers, such as co-acting, similar genres and who-directs-whom. An analysis requirement may also want to use \textit{preferences} for community interactions. As an example, one may be interested in groups (or communities) where the \textit{most important} actors and directors interact. % (Query 4c),
%%while another analysis may seek groups that have maximum interaction between actor and director groups.
%%Similar considerations need to be taken into account when strongly connected community combinations of actors, directors need to be composed with movie ratings.
%Traditional approaches of a) treating the heterogeneous network as a whole by removing type information and applying community detection or b). projecting one layer onto another and then finding communities, will give results that  either hide the effect of different entity types and/or different intra- or inter-layer relationship combinations leading to loss of or misleading information.

%\sharma{2/15/2019}{need to figure out how to phrase this without divulging author identity! rewrote}
%%The community definition and detection research in the literature for homogeneous MLNs~\cite{BDA/SantraB17,ICCS/SantraBC17} are not applicable to HeMLN as each layer has \textit{different sets and types of entities} with  \textit{inter-layer edges} between them.
%It is important to note that this formulation of communities preserves entity and feature types as compared to other alternatives proposed in the literature.
%%We propose a general community definition for HeMLN consisting of any number of layers and arbitrary inter-layer connections.
%We have The problem escalates when new entities and/or relationships get involved like movies having similar IMDB ratings/gross collections/language, production houses, country of release, actor nationality and so on, as the scope of queries increases.
Our \textit{definition of community for a HeMLN uses coupling of communities based on the connection strength (expressed as a weight) and is consistent with the simple graph definition of a community. Further, it also preserves the structure and semantics due to composition which is also shown to be efficient!}. Table~\ref{table:notations} lists all notations used in the paper and their meaning for quick reference.

\subsection{Formal Definition of Community in a HeMLN}
%%\footnote{All the definitions given below are applicable to both a HoMLN and a HeMLN, as well as for disjoint or overlapping communities. However, we are focusing only on disjoint communities in HeMLNs in this paper.}} %remainder of the paper.}}
\label{sec:definition}

%\abhi{2/22/2019}{Shall we say now or later that in this paper we use non-overlapping communities? Plus, bipartite matching will also return non-overlapping matches}
%\abhi{2/22/2019}{Is it an empty inter-layer edge set or bipartite graph?}
%\sharma{2/23/19}{i do not think it is needed for the definition, unless it does not work for them. we can say it in the experimental section if that is what we are using}
%For all of our definitions of a community in a MLN below, there is no need to differentiate between disjoint and overlapping communities.

A \textbf{\textit{1-community}} is a set of communities of the simple graph corresponding to  a layer.
%%A 1-community is represented as a set of singleton tuples each with a community id from that graph (or layer) and an empty inter-layer edge set ($\phi$.)

A \textbf{community bipartite graph\footnote{We defined the set X of bipartite graphs between layers of HeMLN in Section~\ref{sec:problem-statement}. This is a different bipartite graph between two sets of nodes (termed meta nodes) from two distinct layers that correspond to communities in each layer. A single bipartite edge (termed meta edge) is drawn between distinct meta node pairs as defined.}}
\textbf{\textit{CBG$_{i,j}$($U_i$, $U_j$, $L'_{i,j}$)}}
is defined between two disjoint and independent sets $U_i$ and $U_j$. An element of $U_i$ ($U_j$) is a 1-community id from $G_i$ ($G_j$) that is represented as a single meta node. $L'_{i,j}$ is the set of meta edges between the nodes of $U_i$ and $U_j$ (or bipartite graph edges.) For any two meta nodes, a \textit{single edge} is included in $L'_{i,j}$, if there is \textit{an inter-layer edge} between any pair of nodes from the corresponding communities (acting as meta nodes in CBG) in layers $G_i$ and $G_j$. Note that there may be many inter-layer edges between the communities from the two layers. Also note that $U_i$ ($U_j$) need not include all community ids of $G_i$ ($G_j$.) The strength (or weight) component of the meta edges is elaborated in Section \ref{sec:customize-maxflow}.

A \textbf{\textit{serial 2-community}} is defined on the community bipartite graph \textit{CBG$_{i,j}$($U_i$, $U_j$, $L'_{i,j}$)} corresponding to layers $G_i$ and $G_j$
%%\footnote{All 1-community elements from $G_i$ ($G_j$) need not be in $U_i$ ($U_j$.)}.
A 2-community is a set of tuples each with a pair of elements $<c_i^m, c_j^n>$, where $c_i^m \in U_i$ and $c_j^n \in U_j$, that satisfy the \textit{Maximum Weighted Bipartite Coupling (MWBC)} (composition function $\Theta$ defined in Section~\ref{sec:problem-statement}) for the bipartite graph of $U_i$ and $U_j$, along with the set of inter-layer edges between them. The pairing is done from left-to-right (hence it is \textit{not commutative}) and a single community from the left layer can pair with \textit{zero or several communities} from the right layer. That is, \textit{one-to-many or many-to-one pairings} are possible. The lower bound on the number of 2-community is $|U_i| - ~number ~of ~U_i ~nodes ~that ~have ~no ~outgoing ~edges.$

A \textbf{\textit{serial k-community}}\footnote{k represents the number of layers used for computing the community, not the number of compositions. The \textbf{``serial"} prefix used for defining a k-community indicates the order used (but can be arbitrary) in its specification. A k-community corresponds to a connected subgraph of k layers. Our definition assumes left-to-right precedence for the composition function $\Theta$. It is possible to define a k-community with explicit precedence specification for $\Theta$.
Also, other definitions are possible that may be order agnostic. Finally, we drop the repetitive ``serial" prefix henceforth  as we only refer to a serial k-community in the rest of the paper.}  for \textit{k} layers of a HeMLN is defined as the application of \textit{serial 2-community} definition recursively to compose a k-community.
%%\textcolor{blue}{SB: I think the difference between update and extend can be made more clear. Also holds for Table 2 Can we add figures or examples based on IMDB ?}
%%sc; have tried to clarify
The base case corresponds to applying the definition of 2-community for any two layers. The recursive case corresponds to applying  2-community composition for a  t-community with another $G_j$.

For each recursive step, there are two cases for the 2-community under consideration: i) the $U_i$  is from a layer $G_i$ \textit{already in the t-community} and the $U_j$  is from a \textit{new layer $G_j$}. This bipartite graph match is said to \textbf{extend} a t-community (t $<$ k) to a \textit{(t+1)-community},
%%when $U_i$ corresponds to \textit{some layer $G_i$ of the \textit{t-community}} and $U_j$ is a \textit{new layer} $G_j$
or ii) \textbf{both} $U_i$ ($U_j$)  from layers $G_i$ ($G_J$) are \textit{already in the t-community}.
This bipartite graph match is said to \textbf{update} a t-community (t $<$ k), \textbf{not extend} it.

%%%Only the inter-layer edges corresponding to the meta edge \textit{me} between the meta nodes (\textit{expanded(me)} represented as \textbf{$x_{i,j}$} ) is kept for structure preservation.
%%added to the representation to capture the inter-layer edges for this composition.

In both cases i) and ii) above, a number of outcomes are possible. Either a meta node from $U_i$: a) matches one or more meta nodes in $U_j$ resulting in a (or many) \textbf{consistent match}, or b) does not match a meta node in $U_j$ resulting in a \textbf{no match}, or c) matches a node in $U_j$ that is not consistent with a previous match termed \textbf{inconsistent match}.

Structure preservation is accomplished by retaining, for each tuple of t-community,  either a matching community id (or 0 if no match) and
$x_{i,j}$ (or $\phi$ for empty set) representing inter-layer edges corresponding to the meta edge between the meta nodes (termed \textbf{expanded(meta edge).)} The \textit{extend} and \textit{update} carried out for each of the outcomes on the representation is listed in Table~\ref{table:algo-refer}. Note that due to multiple pairing of nodes during any composition, the number of tuples (or t-communities) may increase. Copying is used to deal with multiple pairings.

%%\textbf{Note that a null community id is represented as 0 and a empty expanded meta edge is represented as $\phi$.}

%%it \textbf{updates} a t-community ($t ~<=~ k$) when both $U_i$ and $U_j$  correspond to some layer in the t-community.
%%For both, a community bipartite graph CBG  used for applying the maximum flow matching.

%%The $CBG_{i,j}$($U_i$, $U_j$, $L'_{i,j}$) is used for both. However, for i), $U_i$ corresponds to a \textit{layer $G_i$ of the \textit{t-community}} and $U_j$ a new layer $G_j$. Only the meta nodes that are part of the t-community from layer $G_i$ are used for $U_i$. $U_j$ has all the 1-community nodes from $G_j$ as meta nodes. Whereas for case ii), both $U_i$ and $U_j$ correspond to some layers $G_i$ and $G_j$ of the t-community. For both $U_i$ ($U_j$) only the meta nodes that are part of the t-community from layer $G_i$ ($G_j$) are used.

\begin{table}[h!t]
\centering
%\vspace{-5pt}
    \begin{tabular}{|p{3.6cm}|p{3.9cm}|}
            \hline
                \textbf{($G_{left}$, $G_{right}$) outcome} & \textbf{Effect on tuple \textit{t} } \\
            \hline
            \hline
            \multicolumn{2}{|c|}{\texttt{\textcolor{blue}{case (i) - one processed and one new layer}}} \\
            \hline
            a) \texttt{\textcolor{blue}{consistent match}} & \textbf{Copy \& Extend} \textit{t} with paired community id \textbf{and} $x_{i,j}$\\% (meta edge expanded as an edge set)\\
            \hline
            b) \texttt{\textcolor{blue}{no match}} & \textbf{Copy \& Extend} \textit{t} with 0 and $\phi$\\
            \hline
            \hline
            \multicolumn{2}{|c|}{\texttt{\textcolor{blue}{case (ii) - both are processed layers}}} \\
            \hline
            a) \texttt{\textcolor{blue}{consistent match}} & \textbf{Copy \& Update} \textit{t} \textbf{only with} $x$\\
            \hline
            b) \texttt{\textcolor{blue}{no match}} & \textbf{Copy \& Update} \textit{t} \textbf{only with} $\phi$\\
            \hline
            c) \texttt{\textcolor{blue}{inconsistent match}} & \textbf{Copy \& Update} \textit{t} \textbf{only with} $\phi$\\
            \hline

        \end{tabular}
            \caption{Cases and outcomes for MWBC (Algorithm \ref{alg:k-community})}
%            \vspace{-15pt}
    \label{table:algo-refer}
\end{table}

A HeMLN can be viewed as a simple graph (termed HeMLN-graph) with each layer of a HeMLN being a node and the presence of inter-layer edges between layers denoted by an edge between corresponding nodes. Then, a k-community can be specified over any connected subgraph of the HeMLN-graph. Case i) above corresponds to a k-community of \textit{an acyclic} subgraph of HeMLN-graph and case ii) to a k-community of a  \textit{cyclic} subgraph of the HeMLN-graph. For both, the number of compositions will be determined by the number of edges in the connected subgraph and can be more than the number of layers (or nodes).  Also, for both cases, MWBC algorithm results in one of the 3 outcomes: a \textit{consistent match }, \textit{no match}, or an \textit{inconsistent match (only for case (ii)} as shown in Table~\ref{table:algo-refer}.
%%The t-community is \textbf{updated}. It still remains a \textit{t-community},

%%For i) without loss of generality, let $U_i$ correspond to \textit{layer $G_i$ of the \textit{t-community} (of some t layers)} and $U_j$ a new layer $G_j$. Only the meta nodes that are part of the t-community from layer $G_i$ are used for $U_i$. $U_j$ has all the 1-community nodes from $G_j$ as meta nodes. $L'_{i,j}$, as defined using the MLN, is used and maximum flow match is applied on CBG$_{i,j}$($U_i$, $U_j$, $L'_{i,j}))$.  The t-community is extended to a (t+1)-community on both the components of some tuples. For ii) it is similar to i) except that both $U_i$ and $U_j$ correspond to some layer $G_i$ and $G_j$ of the t-community.
%For both $U_i$ ($U_j$) only the meta nodes that are part of the t-community from layer $G_i$ ($G_j$) are used. Again,  $L'_{i,j}$ is used and the maximum flow mach is applied on CBG$_{i,j}$($U_i$, $V_j$, $L'_{i,j})$. Here, instead of extending, the t-community is updated only on the second component of some tuples based on the matches obtained. Details of extension and update are elaborated in Section~\ref{sec:rep} on the representation. \textcolor{blue}{SB: I think picture would help in understanding the deifnitions better. Also need to be more tightened. We also need several subpargaraphs in the section for ease of reading}

\subsection{Characteristics of k-community}
%%for some nodes in $U_i$  in which case the pairing 2-community is represented by an empty community (denoted by 0) and the  bipartite graph as an empty set (denoted by $\phi$). For case ii), the maximal flow match may result in either \textit{consistent matches} with respect to nodes in $U_j$, or \textit{no matches} for nodes in $U_i$,  or even \textit{inconsistent matches} with respect to nodes in $U_j$
%In both cases, a $\phi$ is added as part of revision.
%Details of extension and revision are discussed as part of the representation and algorithm.

%\abhi{2/14/2019}{Discuss coupling based definition}
%Note that each element of a k-community is an ordering of 1-community from at most k layers  and at most k-1 bipartite graphs from each participating layer.

%Each tuple of a k-community is a sequence of community ids from layers indicating the order of community creation separated by comma followed by a sequence of corresponding bipartite graphs separated by comma. Each recursive application of the definition adds a community and a bipartite graph.
%In other words, the number bipartite graphs in a k-community is equal to the number of distinct edges of the connected subgraph used to define the k-community with k layers. The number of community ids in a k-community is equal to the number of layers, i.e. k.%number of edges in the connected subgraph + 1.
%This definition defines a  k-community for both cyclic and acyclic graphs with k nodes (as layers.)

\begin{figure}[ht]
   \centering
%   \vspace{-15pt}
   \includegraphics[width=\linewidth]{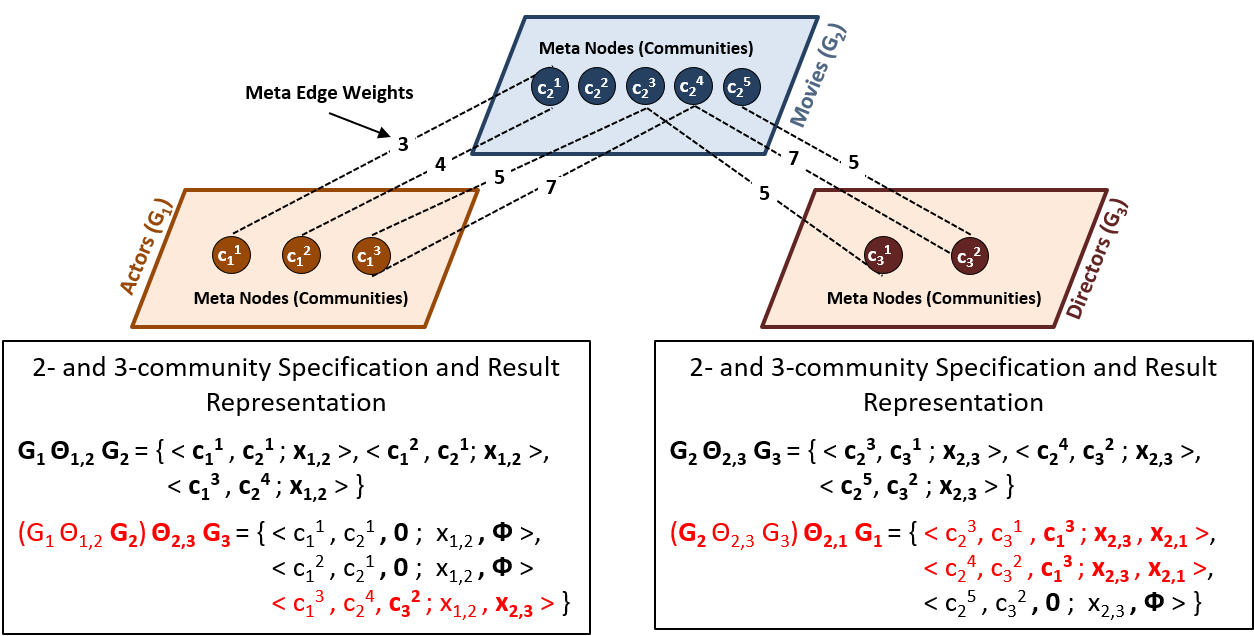}
   \caption{{Illustration of order dependence on a k-community}}
   \label{fig:sequence}
%\vspace{-15pt}
\end{figure}

The above definition when applied to a specification (such as the one shown in Figure~\ref{fig:sequence}) generates \textit{progressively strong coupling between layers} (due to left-to-right precedence of $\Theta$) using MWBC. \textit{Thus, our definition of a k-community is characterized by dense connectivity within the layer (community definition) and strong coupling across layers (MWBC semantics.)} %characterize this definition of k-community which, we believe, matches the original intuition of a community for a simple graph.
%%Note that the traditional methods of finding communities in HeMLN, such as aggregation or projection can be expressed as special cases of our structure-preserving definition.
%Each community of the resulting k-community can be aggregated (or collapsed) into a single type-independent graph or community. It can also be aggregated (or collapsed) using projection as well. Hence, the above definition is structure-preserving, and flexible for independent usage as well as aggregation.

%However, a k-community is meaningfully defined not only in terms of the k layers involved but also using an acyclic graph of those layers.
%\footnote{It is possible to define k-community for cyclic graphs where a layer needs to be used more than once. Since a k-community has k layers and (k-1) bipartite graphs if all layers are distinct, k-community on cyclic graphs will have k layers and k bipartite graphs.}.
%%\textcolor{blue}{SB: Can we bring figure 2 closer so that it is side by side with the text. I dont think we have defined cyclic and acylcic communities before}

{\bf Space of Analysis Alternatives: } Given a HeMLN with k layers, the number of possible k-community (or analysis space) is quite large. For a HeMLN-graph, the number of potential k-community is a function of the number of unique connected subgraphs of different sizes and the number of possible orderings for each such connected subgraph. With the inclusion of 3 weight metrics (see Section~\ref{sec:customize-maxflow}), it gets even larger. It is important to understand that each subgraph of a given size (equal to the number of edges in the connected subgraph) along with the ordering represents a \textit{different} analysis of the data set and provides a different perspective thereby supporting a large space of analysis alternatives. Finally, the composition function $\Theta$ defined above is not commutative (due to left-to-right pairing) and also not associative\footnote{Due to the use of a subset of meta nodes rather than the entire 1-community during any recursive step.}. Hence, for each k-community, the \textit{order in which a k-community} is defined has a bearing on the result (semantics) obtained. In fact, the ordering is important as it differentiates one analysis from the other even for the same set of layers  and inter-layer connections as elaborated in Section~\ref{sec:application-and-analysis}. Figure~\ref{fig:sequence} shows clearly two 3-community for the same layers which are quite different!

{\bf Need for a new pairing algorithm:} In a traditional bipartite graph (used for dating, hiring etc.), each node is a simple node.
%The goal is to find best matches for the smaller set of nodes. Hence, each node from smaller set is paired only with at most one node from the other set.
The goal is to find maximum number of matches (bipartite edges) such that no two matches share the same node. Hence, a node from one set is paired with at most one node from the other set.
This has been extended to include  weights for the edges without changing the  pairing semantics~\cite{edmonds1965maximum}.  On the other hand, for maximal network flow algorithms~\cite{ford_fulkerson_1956}, a source and a sink is assumed and weights have to be given from source to each node which is impractical in our case.

In contrast,  each meta node in our case is a community representing a group of entities with additional characteristics. For a k-community to be meaningful, we need to associate weights with edges to capture not only the number of edges but also characteristics of the participating communities as well.
%%Hence, the weights associated with the edges of the community bipartite graph need to reflect the participating community characteristics which has a significant bearing on the strength of the coupling.
To capture this, we discuss a number of alternatives for weights (termed weight metrics $\omega$) in Section~\ref{sec:customize-maxflow}, derived from real-world scenarios.

For pairing nodes of the bipartite graph, since traditional approaches are not suited for our coupling, we propose a edge weight-based coupling which reflects the semantics of the community. Each node from the first set is paired with \textit{zero, one or more nodes} from the second set solely based on the outgoing edge weights of that node. This is repeated for each node from the first set.
\textit{Most importantly, unlike current alternatives in the literature for community of a MLN, there is no information loss or distortion or hiding the effect of different entity types or relationships in our definition.}

\section{H\MakeLowercase{e}MLN \MakeLowercase{k}-Community Detection}
\label{sec:community-detection}

%Although the definition of a k-community is recursive, it is easier to develop an iterative algorithm for computing a k-community as the order of composition needs to be specified in some way.

In this section, we first present a specification of a k-community and elaborate on a structure-preserving representation for the result. Then we present an algorithm for the maximum weighted bipartite coupling approach discussed earlier. We propose a number of meaningful ways in which we can consider the strength of the coupling for the MWBC approach by providing alternative weight metrics based on participating community characteristics to match with analysis objectives.

%%The serial k-community set, as per definition in Section \ref{sec:hemln-community}, is recursively generated by using the composition function, $\Theta$,  to find the 2-community set between some layer $G_i$ (from (k-1)-community) and layer $G_k$. In order to produce \textit{strong community couplings/pairings} between two layers, in this paper we propose the customization of the bipartite graph matching technique, that maximizes the information flow, for community detection. However, the community matches may not be unique. Based on the analysis requirements, the pairing of communities will depend on the \textit{proportion} of inter-layer interaction between communities from two layers, presence of interaction among community members that \textit{satisfy some graph property like high centrality}, community size, community density and so on. This section introduces representation, composition algorithms and the various metrics that have been used for community detection in order to adhere to the analysis requirements.

% meta node has its own characteristics and that will effect the weight of the edge. draw relations from couple matching. intuition behind giving weights. different from traditional max flow. node semantics into the flow

% HemLN is a forest. each analysis requires a connected subgraph. and then detect communities

\subsection{HeMLN k-Community Representation}
\label{sec:rep}

%%\textcolor{blue}{SB: Is it necessary to write ($G_2$ $\Theta_{2,1}$ $G_1$)  $\Theta_{2,3}$ $G_3$. Since the order of the networks gives the order in which they are combined we could write ($G_2$ $\Theta$ $G_1$)  $\Theta$ $G_3$. Also it might be easier to explain the tuples in the k-communities using an example since the abstraction might be hard to follow. }

%%Essentially, we need to specify a k-community for computation and a representation for storing computed results.
Linearization of a HeMLN structure is done using an order of specification which is also used for computation. Although a k-community need to be specified as an expression involving $\Psi$ and $\Theta$, as indicated earlier, we drop $\Psi$ for clarity. For the layers shown in Figure~\ref{fig:sequence}, a 3-community computation shown is for the specification (($G_2$ $\Theta_{2,3}$ $G_3$)  $\Theta_{2,1}$ $G_1)$. We can drop the parentheses as the precedence of $\Theta$ is assumed. However, we need the subscripts for $\Theta$ to disambiguate a k-community specification when a composition is done on the layers already used.
%%from the above as the order is assumed to be from left to right and the layers involved are clear from the bipartite graph subscripts\footnote{This is because there is only \textit{one edge} between any two layers in our MLN definition.}.
If the layers $G_1 ~ and ~G_3$ were also connected with inter-layer edges in Figure~\ref{fig:sequence}, a 3-community involving a cycle can be specified as $G_1$ $\Theta_{1,2}$ $G_2$  $\Theta_{2,3}$ $G_3 ~\Theta_{3,1}~ G_1$.

A k-community is represented as a set of tuples. Each tuple represents a distinct element of a k-community and includes an ordering of k community ids as items  (a path, if you will, connecting community ids from different layers) and at least (k-1)  expanded(meta edge) (i.e., $x_{i,j}$ elements.) This representation completely preserves the MLN structure along with semantics (labels) to reconstruct a HeMLN for any k-community. It is possible that there are multiple paths originating from the communities in the first layer of the expression due to one to many pairings. That is, a community in a layer can participate in more than one k-community tuple. All these paths need not remain total as the k-community computations progresses\footnote{This is in contrast to the traditional pairing algorithms where any community can participate in \textit{only one path of a k-community}.}. In summary, each k-community is a tuple with 2 distinct components. The first component is a comma-separated ordering of community ids (as items) from a distinct layer. The second component is  also a comma-separated ordering of at least (k-1) $x_{i,j}$ (with each x having a different pair of subscripts.)  Communities for $x$ are uniquely identifiable from the subscripts. Figure~\ref{fig:sequence} shows a number of 2- and 3-community results for the corresponding specifications.
For the acyclic 3-community specification $G_1$ $\Theta_{1,2}$ $G_2$ $\Theta_{2,3}$ $G_3$, the element $<$ $c_1^3$, $c_2^4$, $c_3^2$ \textbf{;} $x_{1,2}$, $x_{2,3}$ $>$ is the only total element as it does not include any $\phi$, whereas, the other two elements, $<$ $c_1^1$, $c_2^1$, 0 \textbf{;} $x_{1,2}$, $\phi$ $>$ and $<$ $c_1^2$, $c_2^1$, 0 \textbf{;} $x_{1,2}$, $\phi$ $>$, are partial as both include one $\phi$. Moreover, the partial elements share the $c_2^1$ showing that multiple paths can pass through the same meta node (community).
%It also shows results where there is a \textit{no match} for maximal network flow for a 3-community.

If $G_1$ and $G_3$ were connected, then for the 3-community specification $G_1$ $\Theta_{1,2}$ $G_2$ $\Theta_{2,3}$ $G_3$ $\Theta_{3,1}$ $G_1$ (involving a cycle), the total element shown in figure changes to \{
% $<$ $c_1^1$, $c_2^3$, 0 \textbf{;} $x_{1,2}$, $\phi$ $>$, $<$ $c_1^2$, $c_2^1$, $c_3^2$ \textbf{;} $x_{1,2}$, $x_{2,3}$, $x_{3,1}$ $>$, $<$ $c_1^3$, $c_2^5$, 0 \textbf{;} $x_{1,2}$, $\phi$ $>$
%$<$ $c_2^4$, $c_3^1$, $c_1^1$ \textbf{;} $x_{2,3}$, $x_{2,1}$, $x_{1,3}$ $>$, $<c_2^1$, $c_3^2$, $c_1^2$ \textbf{;}  $x_{2,3}$, $x_{2,1}$, $\phi$ $>$
$<$ $c_1^3$, $c_2^4$, $c_3^2$ \textbf{;} $x_{1,2}$, $x_{2,3}$, $x_{3,1}$ $>$
\}. In this case, the number of communities in each tuple is \textit{k} (3 here) and the number of inter-layer edge sets is \textit{at least (k-1)}.
%%It is exactly (k-1) if the k-community is for an acyclic connected graph and more depending upon the number of edges in cyclic subgraph (here it is 3 as one cyclic edge is included.)
To generalize, an element of \textit{\textbf{k-community}} for an arbitrary specification \\
$G_{n1}$ $\Theta_{n1,n2}$ $G_{n2}$ $\Theta_{n2,n3}$ $G_{n3}$ ... $\Theta_{ni,nk}$ $G_{k}$ \\ will be represented as \\ $<$ $c_{n1}^{m1}$, $c_{n2}^{m2}$, ..., $c_{nk}^{mk}$ \textbf{;} $x_{n1,n2}$, $x_{n2,n3}$, ..., $x_{ni,nk}$ $>$, where some c's may be 0 and some x's may be $\phi$.
\subsection{Structure-Preserving Representation Benefits}

\begin{enumerate}
    \item Each element of a k-community can be further analyzed individually as the tuple contains all the information to reconstruct the HeMLN and drill down for details.
    \item Total and partial elements of k-community provide important information about the result characteristics.  A partial community shows a weak coupling of the complete community whereas a total element indicates strong coupling.
    \item The resulting set can be ranked in several ways based on HeMLN community characteristics. For example, they can be ranked based on community size or density (or any other feature) as well as significance of the layer.
%%\item Finally, results from previous specifications can be re-used if the same sub-expression needs to be evaluated again.

\end{enumerate}
\subsection{HeMLN k-Community Detection Algorithm}
\label{sec:flow}

The MWBC algorithm identifies pairs of communities for the community bipartite graph input along with edge weights. Each node from the left set is paired with zero or more nodes from the other set. Either the highest edge weight pairs or all pairs with equal weight are output. This is important as the coupling strength is the same with multiple communities and all of them need to be in the result. This algorithm has been implemented using a single pass of the bipartite graph edges. Note that the number of communities from each layer is likely to be significantly less than the number of entity nodes in that layer. This translates to  significantly less number of meta edges in the bipartite graph compared to the total number of inter-layer edges between the layers.

\begin{algorithm}[h]
\caption{HeMLN k-community Detection Algorithm}
\label{alg:k-community}
\begin{algorithmic}[1]
\REQUIRE- \\
   \textbf{INPUT:}  HeMLN, ($G_{n1}$ $\Theta_{n1,n2}$ $G_{n2}$ ... $\Theta_{ni,nk}$ $G_{nk}$), and a weight metric (wm). \\

   \textbf{OUTPUT:} Set of distinct k-community tuples\\
    %%\texttt{//input processed from left to right} \\
    \STATE \textbf{Initialize:} k=2, $U_i$ = $\phi$, $U_j$ = $\phi$,  result$'$ = $\emptyset$\\
    \textit{result} $\gets$ MWBC($G_{n1}$,$G_{n2}$, HeMLN, wm) \\
    \textit{left}, \textit{right} $\gets$ left and right subscripts of second $\Theta$ \\

    \WHILE{\textit{left} $\neq$ null \&\& \textit{right} $\neq$ null}
        %\texttt{// subset if the layer has been processed} \\
        \STATE $U_i$ $\gets$ subset of 1-community($G_{left}, result$) \\
        \STATE $U_j$ $\gets$ subset of     1-community($G_{right}, result$) \\
        %%\texttt{// s pubsets (3,4) if layer has been processed} \\

        \STATE \textit{MP} $\gets$ MWBC($U_i$, $U_j$, HeMLN, wm) \\
        \texttt{//a set of comm pairs $<c_{left}^p$,$c_{right}^q>$} \\
        \FOR{ \textbf{each} tuple \textit{t} $\in$ \textit{result} } \STATE kflag = false
            \IF {\textit{both $c_{left}^x ~and~ c_{right}^y$} are part of \textit{t} and $\in$ MP \texttt{\textcolor{blue}{[case ii (processed layer): consistent match]}}}
                \STATE Update \textit{a copy of t} with  ($x_{left, ~right}$) and append to result$'$
            \ELSIF {$c_{left}^x$ is  part of \textit{t} and $\in$ MP and $G_{right}$ layer has been processed \texttt{\textcolor{blue}{[case ii (processed layer): no and inconsistent match]}} }
                \STATE Update \textit{a copy of t} with $\phi$ and append to result$'$\\
            \ELSIF {$c_{left}^x$ is part of \textit{t} and \textcolor{red}{for each} $c_{left}^x ~\in~ MP$  \texttt{\textcolor{blue}{[case i (new layer): consistent match]}}}
                \STATE copy and Extend \textit{t} with paired $c_{right}^y$ $\in$ MP and $x_{left, ~right}$ and append to result$'$;  kflag = true \\
            \ELSIF {$c_{left}^x$ is part of \textit{t} and $\notin$ MP \texttt{\textcolor{blue}{[case i (new layer): no match]}}}
                \STATE copy and Extend \textit{t} with 0 (community id) and $\phi$ and append to result$'$; kflag = true \\
            \ENDIF
        \ENDFOR \\
        \textit{left}, \textit{right} = next left, right subscripts of $\Theta$ or null\\
        if kflag k = k + 1; result = result$'$; result$'$ = $\emptyset$
        \ENDWHILE
     \end{algorithmic}
     %\vspace{-15pt}
\end{algorithm}

%%\abhi{2/28/2019}{Check the sentence for k-communities. I have revised it ad commented the earlier sentence.}
Algorithm~\ref{alg:k-community}  accepts a linearized specification of a k-community and computes the result as described earlier. The input is an \textit{ordering of layers}, \textit{composition function indicating the community bipartite graphs to be used} and the type of weight to be used. %The output is a \textit{set} whose size indicates the number of distinct k-communities for that specification (which is bound by the base case) and \textit{each element of that set is a tuple corresponding to a distinct, single k-community}.
The output is a \textit{set} whose \textit{elements are tuples corresponding to  distinct, single HeMLN k-community} for that specification. The size (i.e., number of tuples) of this set is determined by the pairs obtained during computation. The layers for any 2-community bipartite graph composition are identifiable from the input specification. %%he result, as described earlier, preserves the structure and all intra- and inter-related edge information for its reconstruction.

%\abhi{2/25/2019}{Isn't number of iterations of while loop: number of theta - 1?}
The algorithm iterates until there are no more compositions to be applied. The number of iterations is equal to the number of $\Theta$ in the input (corresponds to the number of inter-layer connections.) For each layer, we assume that its 1-community has been computed.
%%This can be done earlier or on demand, and sequentially or in parallel. Communities are assumed to be uniquely identified and includes layer id.

The bipartite graph for the base case and for each iteration is constructed for the participating layers (either one is new or both are from the t-community for some t) and MWBC algorithm is applied. The result obtained is used to either extend or update the tuples of the t-community and add new tuples as well. All cases are described in Table~\ref{table:algo-refer}.
Note that the k-community size\textbf{ k is incremented} only when a \textit{new layer is composed (case i).)} For case ii) (cyclic k-community) \textbf{k is not incremented} when \textit{both layers are part of the t-community}.
When the algorithm terminates, we will have the set of tuples each corresponding to a single, distinct  k-community for the given specification.

%%It is not necessary to use the \textit{same weight metric} for all bipartite graph matches while computing a k-community. It is possible to choose different metrics for each bipartite match to adhere to the analysis semantics. Algorithm~\ref{alg:k-community}, as given, does not support this.

Figure~\ref{fig:sequence} illustrates examples of 2- and 3-community results computed using the above algorithm. The figure also shows the two components of each tuple comprising of community id as items in the first component and the set of expanded(meta edge) or $x$ as items for the second component. Further, it shows how the result of a 2-community is extended to form a 3-community.  It also demonstrates the importance of order in which a k-community is defined. Further, it shows that the same result need be not obtained for different ordering (e.g., 3-community) \textit{on the same layers}.

\section{Customizing MWBC}
\label{sec:customize-maxflow}

Algorithm~\ref{alg:k-community} in Section~\ref{sec:community-detection} uses a bipartite graph match with a given weight metric. As we indicated earlier, there is an important difference between simple and \textbf{meta nodes/edges} that represent a \textbf{community of nodes/set of edges}. Without including the characteristics of meta nodes and edges for the match, we cannot argue that the pairing obtained represents analysis based on  participating community characteristics. Hence, it is important to identify how qualitative community characteristics can be mapped quantitatively to a weight metric (that is, weight of the meta edge in a community bipartite graph) to influence the bipartite matching. Below, we propose three weight metrics and their intuition.

%%All quantitative edge weights are normalized. They include: i) the number of \textit{actual edges} between each pair of  communities in the bipartite graph, ii) \textit{density and edges} between participating communities, and iii) for the nodes in the participating community, \textit{the popularity of \textit{individual nodes} (i.e., hub)} with the nodes in the connected community. i) is useful for characterizing higher interaction between the communities as  requested in \ref{list:maximum-interaction}. Similarly, ii) includes the strength of the participating communities by including density which can be used for ~\ref{list:density}, and finally, iii) captures strong connectivity across central nodes in participating communities using hubs which is appropriate for ~\ref{list:hub-participation}. The criticality of ordering specification in a k-community is exemplified by ~\ref{list:We-acyclic3-adm},  ~\ref{list:We-acyclic3-amd}. Finally, ~\ref{list:We-cyclic3-adma} highlights a query that requires cyclic 3-community. The specifications shown can be matched with the analysis requirements based on the discussions so far.

\subsection{Number of Inter-Community Edges ($\omega_e$)}
\label{sec:m1}
%Certain queries will need the detection of groups where apart from strong interaction among the same type of entities, \textit{maximum interaction exists among entities from different sets}. Therefore, in this case, weight of the inter-layer meta edges should reflect number of interactions that exist between the members of every community pair from two layers.  Formally,

This metric uses number of inter-community edges of participating communities as weight (normalized) for meta-edges. The intuition behind this metric is \textit{maximum connectivity} (size of the community is to some extent factored into it) without including other community characteristics. This weight connotes \textit{maximum interaction between two communities}.
%%(used for \ref{analysis:DBLPHe-PAuY} and \ref{analysis:IMdbHe-madm}.) 

\begin{comment} for icdm
For every meta edge $(u_i^m, u_k^n)$ $\in$ $L'_{i,k}$, where $u_i^m$ and $u_k^n$ are the meta nodes corresponding to communities $c_i^m$ and $c_k^n$, respectively, in the community bipartite graph, the weight, 

\textit{\noindent $\omega_e$($u_i^m$, $u_k^n$) =  $\frac{|x_{i,k}|}{max(\omega_e)}$}, ~where \\
%\textit{\noindent where $L_{i,k}^{m,n}$ = $\bigcup ~\{(a, b): {a \in c_i^m, ~  b \in c_k^n}, ~and~ (a, b) \in L_{i,k}\}$}
\textit{\noindent $x_{i,k}$  = $\bigcup~~ \{(a, b): a \in v_i^{c^m}, b \in v_k^{c^n}, ~and~ (a, b) \in L_{i,k}\}$. }
\end{comment}

\subsection{Hub Participation ($\omega_h$)}
\label{sec:m3}

For many analysis, we are interested in knowing whether highly influential nodes within a community also interact across the community. This can be translated to the \textit{participation of influential nodes within and across each participating community} for analysis. This can be modeled by using the notion of  \textbf{hub}
%%\footnote{High centrality nodes (or hubs) have been defined based on different metrics, such as degree centrality (vertex degree), closeness centrality (mean distance of the vertex from other vertices), betweenness centrality (fraction of shortest paths passing through the vertex), and eigenvector centrality.}
\textbf{participation} within a community and their interaction across layers. In this paper, we have used degree centrality for this metric to connote higher influence. Ratio of participating hubs from each community and the edge fraction are multiplied to compute $\omega_h$. Formally,

For every $(u_i^m, u_k^n)$ $\in$ $L'_{i,k}$, where $u_i^m$ and $u_k^n$ are the meta nodes denoting the communities, $c_i^m$ and $c_k^n$ in the community bipartite graph, respectively, the weight,

\textit{\centerline{ $\omega_h$($u_i^m$, $u_k^n$) =  $\frac{|H_{i,k}^{m,n}|}{|H_i^m|}$ * $\frac{|x_{i,k}|}{|v_i^{c^m}|*|v_k^{c^n}|}$*$\frac{|H_{k,i}^{n,m}|}{|H_k^n|}$,}
}
\textit{\noindent where $x_{i,k}$ = $\bigcup~~\{(a, b): a \in v_i^{c^m}, b \in v_k^{c^n}, ~and~ (a, b) \in L_{i,j}\}$; $H_i^m$ and $H_k^n$ are set of hubs in $c_i^m$ and $c_k^n$, respectively; $H_{i,k}^{m,n}$ is the set of hubs from $c_i^m$ that are connected to $c_k^n$; $H_{k,i}^{n,m}$ is the set of hubs from $c_k^n$ that are connected to $c_i^m$ }.

%%From the analysis description, ~\ref{analysis:IMdbHe-da-wh} matches this metric. 
%%In addition to the above, many other useful weight metrics have been explored to meet analysis requirements. Due to space constraints, we have included what we think are the most useful based on our  analysis experience.
%%giving rise to different community pairings or 2-communities like hubs based on other centrality metrics, unique fraction of community-wise nodes participating in the interaction and so on. However, the important aspect to be noted here is that the meta-bipartite graph structure need not be recreated while accommodating the new set of weights. These further justifies the advantages of the proposed decoupled approach as the partial layer-wise results can be combined based on \textit{any type} of requirements without having to re-compute any of the layer-wise results. 

\subsection{Density and Edge Fraction ($\omega_d$)}
\label{sec:m2}

The intuition behind this metric is to bring participating community density which captures internal structure of a community. Clearly, \textit{higher the densities and larger the edge fraction, the stronger is the interaction (or coupling) between two meta nodes (or communities.)} Since each of these three components (each being a fraction) increases the strength of the inter-layer coupling, they are  multiplied to generate the weight of the meta edge.  The domain of this weight will be $(0,1]$. The weight computation formula is similar to the previous one. 
%%~\ref{analysis:DBLPHe-PAu-wd} uses this weight metric based on the analysis description.

\begin{comment} for ICDM
For every $(u_i^m, u_k^n)$ $\in$ $L'_{i,j}$, where $u_i^m$ and $u_k^n$ denote the communities, $c_i^m$ and $c_k^n$ in the community bipartite graph, respectively, the weight,

\textit{\centerline{$\omega_d$($u_i^m$, $u_k^n$) =  $\frac{2*|e_i^{c^m}|}{|v_i^{c^m}|*(|v_i^{c^m}|-1)}$ * $\frac{|x_{i,k}|}{|v_i^{c^m}|*|v_k^{c^n}|}$*$\frac{2*|e_k^{c^n}|}{|v_k^{c^n}|*(|v_k^{c^n}|-1)}$,}
}
%\textit{\noindent where $L_{i,k}^{m,n}$ = $\bigcup~~ \{(a, b): a \in c_i^m, b \in c_k^n, ~and~ (a, b) \in L_{i,k}\}$}
\textit{\noindent where $x_{i,k}$ = $\bigcup~~ \{(a, b): a \in v_i^{c^m}, b \in v_k^{c^n}, ~and~ (a, b) \in L_{i,k}\}$}
\end{comment}

%%\subsection{Uniqueness of Proposed Metrics}
%%\label{sec:compare-metrics}

Ideally, the alternatives for metrics should be independent of each other so they are useful for different analysis. Also, it is important that their computation be efficient (see Section~\ref{sec:cost-analysis}.) We believe that the three metrics proposed satisfy the above and our experiments have confirmed them although not shown in this paper due to page constraints.

\subsection{HeMLN k-community Detection Efficiency}
\label{sec:cost-analysis}

For a given specification of a k-community, its detection has several cost components. Below, we summarize their individual complexity and cost.

\begin{enumerate}
    \item \textbf{Cost of generating 1-community}: For each layer (or a subset of needed layers) this can  be done in parallel bounding this \textbf{one-time cost} to the largest one (typically for a layer with maximum density.)
    \item \textbf{Cost of computing meta edge weights}: For the proposed analysis metrics, part of them, again, are \textbf{one-time costs} and are calculated independently on the results of 1-community. The  costs for $\omega_d$ and $\omega_h$ require a single pass of the communities  using their node/edge details generated by the community detection algorithm.

    \item \textbf{The recurring cost }(base case and each iteration): This includes the cost of generating the bipartite graph, computing the weight of each meta edge of the community bipartite graph for a given $\omega$, and the MWBC algorithm cost. Only the edge fraction (or the maximum number of edges) and participating hubs need to be computed during each iteration. The cost of MWBC algorithm used in our experiments is O($|E|$), where E is the number of meta edges in the community bipartite graph.  The bipartite graph is generated during the computation of weights for the meta edges. Luckily, in our community bipartite graph, the number of meta edges is \textbf{order of magnitude less} than the number of edges between layers. Also, the number of meta nodes is bound by the number of pairings in the previous iteration.
    
    %% for ICDM \item The components costs within each iteration (and also the base case) are: i) weight computation, community bipartite graph creation, and WBC algorithm cost.  
\end{enumerate}

%%In summary, for the proposed decoupling approach, the bulk of the cost is \textbf{one-time} (1-community detection and portions of $\omega_d ~and~ \omega_h$.) \textbf{Cost within each iteration is insignificant} compared to the one-time cost which is also borne out by the experimental analysis in Section~\ref{sec:experiments}.
\section{Analysis of IMD\MakeLowercase{b} and DBLP Data Sets}
\label{sec:application-and-analysis}

%%In order to demonstrate the need and utility of structure-preserving k-community and identify a set of metrics that are domain-independent and applicable for specific analysis needs, 
We introduce the data sets and analysis objectives to formulate the k-community specification along with the choice of weight metric to apply the algorithm. This will clarify the usage of the three weight metrics for MWBC algorithm. The same data sets are used for experimental analysis.\\

\noindent \textbf{DBLP data set~\cite{data/type2/DBLP}:} The DBLP data set captures information about published research papers in conference/journal, year of publication and the authors. Most readers are familiar with this data set.\\

 %This is a large data set consisting of movie and TV episode data from their beginnings. This data set can be modeled and analyzed in multiple ways.

%%This data set has been chosen for its versatility in that it can be modeled using HoMLN as well as HeMLN based on analysis requirements.\\

%%\noindent \textbf{General Analysis Requirements: }\textit{The analysis goal for this data set is to analyze the three important groups (entity types)  in this data set: actors, directors, and movies based on the information available (attributes.) Given three entity types, we are interested in analyzing directs-actor relationship, acts-in-a-movie with a specific-rating relationship, and directs-movie with a specific review rating relationship.}
%identify strong groups of directors who have directed specific genres and have also directed actors who have acted together. This analysis is extended to include movies that are rated in a particular way for the above groups. 
%%Below, the above analysis is expressed as detailed requirements to facilitate mapping to a k-community. We also indicate the specification for each detailed analysis including the order of composition followed by metric to be used. The rationale for these mappings are discussed in the remainder of the paper. \\

\noindent\textbf{DBLP detailed Analysis Objectives}
	\begin{enumerate}[label={(A\arabic*)}]

%    \item \label{analysis:DBLPHe-PAu-we} Who are the \textit{most popular} collaborators for each conference?
    
%    2-community: P $\Theta_{P,Au}$ Au; $\omega_e$

    \item \label{analysis:DBLPHe-PAu-wd} For each conference, which is the \textit{most cohesive} group of authors who publish frequently?
    
    2-community: P $\Theta_{P,Au}$ Au; $\omega_d$
    
%    \item \label{analysis:DBLPHe-PAu-wh} For each conference, which are the authors groups whose \textit{most collaborative members} publish?
    
%    2-community: P $\Theta_{P,Au}$ Au; $\omega_h$    

%    \item \label{analysis:DBLPHe-PAu-we} Which are the \textit{most preferred} conferences for each author group?
    
%    2-community: Au $\Theta_{P,Au}$ P; $\omega_e$    
    
    \item \label{analysis:DBLPHe-PAuY} For the most popular collaborators from each conference, which are the 3-year period(s) when they were most active? 

    3-community: P $\Theta_{P,Au}$ Au $\Theta_{Au,Y}$ Y; $\omega_e$
    
\end{enumerate}

Based on the DBLP analysis requirements, three layers are modeled for the HeMLN. Layer \textit{Au} connects any two authors (nodes) who have published at least three research papers together. Layer \textit{P} connects research papers (nodes) that appear in the same conference. Layer \textit{Y} connects two year nodes if they belong to same pre-defined period. The inter-layer edges depict \textit{wrote-paper ($L_{Au,P}$), active-in-year ($L_{Au,Y}$) and published-in-year ($L_{P,Y}$)}. For this paper, we have chosen all papers that were published from 2001-2018 in top conferences. Six 3-year periods have been chosen: [2001-2003], [2004-2006], ..., [2016-2018]. \\

\noindent \textbf{IMDb data set~\cite{data/type2/IMDb}: }The IMDb data set captures movies, TV episodes, actor, directors and other related information, such as rating.\\

\noindent\textbf{IMDb Detailed Analysis Objectives}
	\begin{enumerate}[label={(A\arabic*)}, resume]
%	\item \label{analysis:IMdbHe-1} Identify \textit{versatile} director groups who work with \textit{most sought after} actors among co-actors

% REMOVED TO SAVE SPACE
%    \item \label{analysis:IMdbHe-ad} Which similar movie-genre based \textit{groups} of directors and actors should team up such that the \textit{most versatile participating} actors and directors from their corresponding groups have \textit{worked with most of the members}?

    % 2-community: A $\Theta_{A,D}$ D; $\omega_h$; order does not matter.

%	\item \label{analysis:IMdbHe-2} For the group of directors (who direct similar genres) having \textit{maximum interaction} with members of co-actor groups, identify the \textit{most popular rating} for the movies they direct? 
    
%    \item \label{analysis:IMdbHe-adm} Which group of \textit{densely connected} directors (who direct similar genres) also interact more with \textit{densely connected} genre-based actor groups, identify the rating class for the movies they direct? 

    %Acyclic 3-community: A $\Theta_{A,D}$ D $\Theta_{D,M}$ M; $\omega_e$
 
    \item \label{analysis:IMdbHe-da-wh} Based on similarity of genres, for each director group which are the actor groups whose \textit{majority of the most versatile members interact}?
    
    2-community: D $\Theta_{A,D}$ A; $\omega_h$
    
    % \item \label{analysis:IMdbHe-mad} For the \textit{most popular} actor groups from each movie rating class, which are the genre-based director groups with which they have \textit{maximum interaction}?

    %Acyclic 3-community: M $\Theta_{M,A}$ A $\Theta_{A,D}$ D; $\omega_e$
    
%    \item \label{analysis:IMdbHe-amd} For each movie rating class, which are the \textit{most popular} actor and director groups? 

%    Acyclic 3-community: A $\Theta_{A,M}$ M $\Theta_{M,D}$ D; $\omega_e$
    
%	\item \label{analysis:IMdbHe-madm} Find the actor groups with \textit{strong movie ratings} that have \textit{high interaction} with those director groups who also make movies with similar ratings (as that of actor groups)
	% Cyclic 3-community: M $\Theta_{M,A}$ A $\Theta_{A,D}$ D $\Theta_{D,M}$ M;  $\omega_e$

% 	\item \label{analysis:IMdbHe-amda} For movie rating classes, which are the \textit{most popular} actor and director groups that have \textit{strong interaction} among them?
	\item \label{analysis:IMdbHe-madm}For the \textit{most popular} actor groups, for each movie rating class, find the director groups with which they have \textit{maximum interaction} and who also make movies with similar ratings. 
	
	Cyclic 3-community: M $\Theta_{M,A}$ A $\Theta_{A,D}$ D $\Theta_{D,M}$ M;  $\omega_e$
%\label{list:analysis-cyclic3-madm}

% 	Cyclic 3-community: A $\Theta_{A,M}$ M $\Theta_{M,D}$ D $\Theta_{D,A}$ A;  $\omega_e$
	\end{enumerate}

For addressing the IMDb analysis requirements, three layers for the IMDb data set are formed as follows. Layer \textit{A} and Layer \textit{D} connect actors and directors who act-in or direct \textit{similar genres frequently} (intra-layer edges), respectively. Layer \textit{M} connects movies within the same rating range. The inter-layer edges depict \textit{acts-in-a-movie ($L_{A,M}$), directs-movie ($L_{D,M}$) and directs-actor ($L_{A,D}$)}. There are multiple ways of quantifying the similarity of actors and directors based on movie genres they have worked in. A vector was generated with the number of movies for each genre he/she has acted-in/directed. In order to consider the similarity with respect to \textit{frequency of genres}, in layer \textit{Genre}, two actors/directors are connected if the Pearsons' Correlation between their corresponding genre vectors is at least 0.9\footnote{The choice of the coefficient is not arbitrary as it reflects relationship quality. The choice of this value can be based on how actors/directors are weighted against the genres. We have chosen 0.9 for connecting actors in their top genres.}. Moreover, 10 ranges are considered - [0-1), [1-2), ..., [9-10] for movie ratings.

%The inter-layer edges have the following semantics. The directs-actor inter-layer edges ($L_{A,D}$) connect a director with an actor if he has directed that actor in a movie. The  directs-movie inter-layer relationship ($L_{D,M}$) connects a director to his movies in the movie-ratings layer. Finally, acts-in-a-movie relationship is captured by the inter-layer edges ($L_{A,M}$) which connects an actor with the movies the actor has acted in. This exercise also highlights the modeling using a HeMLN for the set of analysis requirements. We have purposely chosen a cyclic HeMLN for illustrative purposes. Figure~\ref{fig:sequence} shows these layers, two inter-layer connections along with some communities in each layer.

For a specific analysis, the characteristics of the communities connected in the bipartite graph need to be used as meta edge weight to get desired coupling. For example, \textit{maximum interaction} and \textit{most popular} in ~\ref{analysis:DBLPHe-PAuY} and ~\ref{analysis:IMdbHe-madm}, are interpreted as the number of edges between the participating communities. In contrast, interaction with cohesive groups as in~\ref{analysis:DBLPHe-PAu-wd}, is interpreted to include community density as well. Versatility is mapped to participation of hub nodes in each group as in  ~\ref{analysis:IMdbHe-da-wh}.

To compute a k-community, k needs to be identified. \ref{analysis:DBLPHe-PAu-wd} and \ref{analysis:IMdbHe-da-wh} require 2-community. ~\ref{analysis:DBLPHe-PAuY} requires 3-community (for 3 layers) with an acyclic specification (using only 2 edges) starting with the Paper layer as the analysis is from that perspective (each Layer P community corresponds to a conference). ~\ref{analysis:IMdbHe-madm} requires a cyclic 3-community using inter-layer relationships between all layers in a particular order. Note that the analysis objectives have been chosen carefully to cover the weights discussed in the paper. The limitation on the number of analysis objectives is purely due to space constraints.

%interaction between the entities of different types can have multiple interpretations. In this paper, broadly we have discussed three classes of interpretations based on aggregation of interaction (Sec \ref{sec:m1}), density of interacting communities (Sec \ref{sec:m2}) and participating hubs in the interaction (Sec \ref{sec:m3}). In this section, we discuss how these different interpretations will effect the weights of the inter-layer meta edges in the meta-bipartite graph such that the maximal flow based matching algorithms give the community pairings with desired characteristics. Moreover, we will show how the requirements of the queries listed in Section \ref{sec:modeling} can be mapped on one of these. 

\section{Experimental Analysis}
\label{sec:experiments}
We would like to point out that the choice of data sets and sizes were mainly for demonstrating the versatility of analysis using the k-community detection and its efficiency as well as drill-down  capability based on structure- and -semantics preservation. We are not trying to demonstrate scalability in this paper. Also, instead of presenting communities, we have chosen to show a few important drill-down results to showcase the structure- and semantics-preservation of our approach.

%%In this section, we present the results of our  analysis of  ~\ref{analysis:DBLPHe-PAu-wd} through ~\ref{analysis:IMdbHe-madm} discussed in Section~\ref{sec:application-and-analysis} on the DBLP~\cite{data/type2/DBLP} and IMDb data set~\cite{data/type2/IMDb} modeled as a HeMLN (as discussed in Section \ref{sec:application-and-analysis}.) %To recap, we have 3 layers,  Actor (layer A), Director (layer D), and Movies (layer M) and 3 inter-layer relationships (directs-actor, acts-in-a-movie, and directs-movies.) We start with the individual layer analysis and  continue with all the 4 analysis using the approach proposed in this paper and discuss their corroboration with the intuitive expectations and analysis discussed earlier.

\noindent \textbf{Experiment Setup:} For DBLP HeMLN, research papers published from 2001-2018 in VLDB, SIGMOD, KDD, ICDM, DaWaK and DASFAA were chosen. For IMDb HeMLN,  we extracted, for the top 500 actors, the movies they have worked in (7500+ movies with 4500+ directors). The actor set was repopulated with the co-actors from these movies, giving a total of 9000+ actors. For this set of actors, directors and movies, the HeMLN with 3 layers described in Section \ref{sec:application-and-analysis} was built.
Widely used Louvain method (\cite{DBLP:Louvain}) was used to detect layer-wise 1-communities. %Infomap~\cite{InfoMap2014}, that works in a hierarchical fashion while optimizing the \textit{map equation}, which exploits the \textit{information - theoretic duality} between the problem of compressing data, and the problem of detecting and extracting significant patterns or structures (communities) within those data based on \textit{information flow}. % and Louvain~\cite{DBLP:Louvain}.
%We use the publicly available package from NetworkX for the bipartite graph match~\cite{Python:maximal-match}.
The k-community detection algorithm \ref{alg:k-community} was implemented in Python version 3.6 and was executed on a quad-core $8^{th}$ generation Intel i5 processor Windows 10 machine with 8 GB RAM.

%\abhi{5/24/2019}{Placeholders have been added for the a) individual layer statistics, b) 4 analysis results (word cloud based figures need to be added.) c) efficiency}

\subsection{Analysis Results}

\noindent\textbf{Individual Layer Statistics}:
Table \ref{table:IMDbHeMLNStats} shows the  layer-wise statistics for IMDb HeMLN. 63 Actor (A) and 61 Director (D) communities based on similar genres are generated. Out of the 10 ranges (communities) in the movie (M) layer, most of the movies were rated in the range [6-7), while least popular rating was [1-2). No movie had a rating in the range [0-1).
\begin{table}[h!t]
\renewcommand{\arraystretch}{1}
\centering
%\vspace{-10pt}
    \begin{tabular}{|c|c|c|c|}
        \hline
        & \textbf{Actor} & \textbf{Director} & \textbf{Movie} \\
        \hline
        \textbf{\#Nodes} & 9485 & 4510 & 7951 \\
        \hline
        \textbf{\#Edges} & 996,527 & 250,845 & 8,777,618  \\
        \hline
        \textbf{\#Communities (Size $>$ 1)} & 63 & 61 & 9 \\
        \hline
        \textbf{Avg. Community Size} & 148.5 & 73 & 883.4 \\
        \hline
        \end{tabular}
\caption{IMDB HeMLN Statistics}
\label{table:IMDbHeMLNStats}
%\vspace{-10pt}

\end{table}

Similarly, DBLP HeMLN statistics are shown in Table \ref{table:DBLPHeMLNStats}. 591 Author (Au) communities are generated based on co-authorship. 6 Paper (P) communities are formed by grouping papers published in same conference. KDD (2942) and DASFAA (583) have highest and least published papers, respectively. Out of 6 ranges of years (Y) selected, the maximum and minimum papers were published in 2016-2018 (1978) and 2001-2003 (1421), respectively.

\begin{table}[h!t]
\renewcommand{\arraystretch}{1}
\centering
%\vspace{-10pt}
    \begin{tabular}{|c|c|c|c|}
        \hline
        & \textbf{Author} & \textbf{Paper} & \textbf{Year} \\
        \hline
        \textbf{\#Nodes} & 16,918 & 10,326 & 18 \\
        \hline
        \textbf{\#Edges} & 2,483 & 12,044,080 & 18 \\
        \hline
        \textbf{\#Communities (size $>1$)} & 591 & 6 & 6 \\
        \hline
        \textbf{Avg. Community Size} & 3.3 & 1721 & 3 \\
        \hline
        \end{tabular}
\caption{DBLP HeMLN Statistics}
\label{table:DBLPHeMLNStats}
%\vspace{-15pt}

\end{table}

 \begin{figure}[h]
   \centering
%   \vspace{-5pt}
   \includegraphics[width=0.7\linewidth]{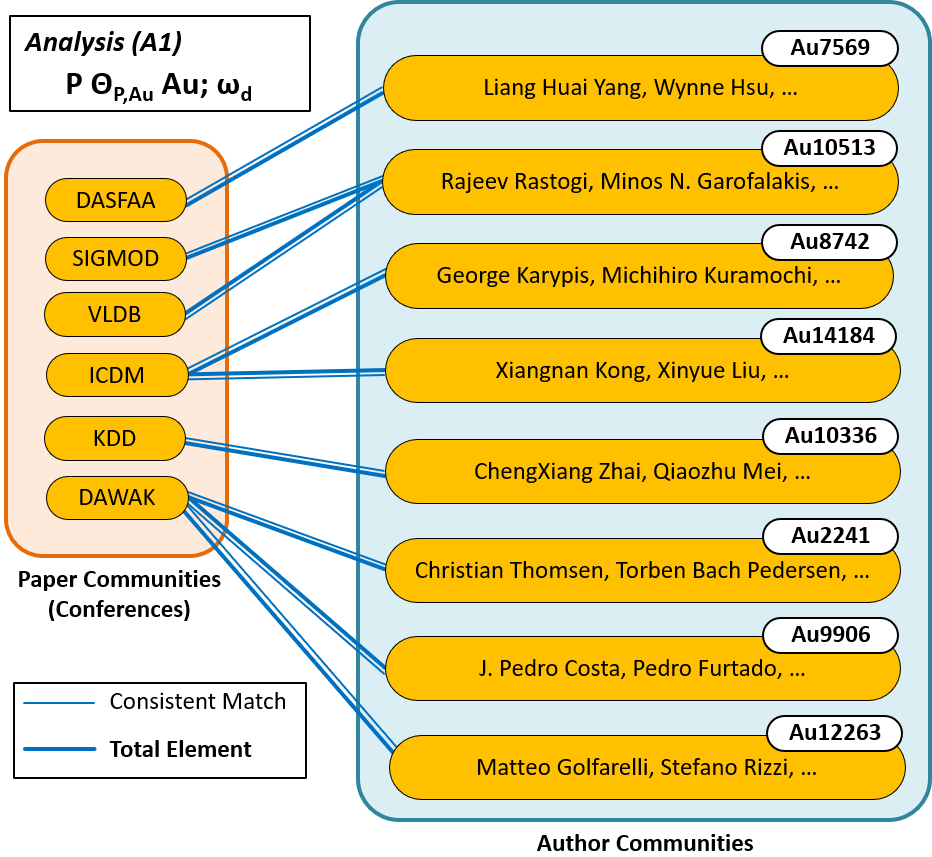}
   \caption{{\ref{analysis:DBLPHe-PAu-wd} Result}: \textbf{7 Total Elements\protect\footnotemark}}
   \label{fig:DBLPHe-PAu-wd}
%\vspace{-10pt}
\end{figure}

\footnotetext{Louvain numbers all communities from 1 and we only consider communities having \textit{at least two members} for this paper. The numbering used in the paper have layer name followed by the Louvain-generated community ID (e.g. A91, Au8742).}

\noindent\textbf{\ref{analysis:DBLPHe-PAu-wd} Analysis: }
%For each of 6 conferences, we obtained most cohesive author (Au) communities that also has high interactions among themselves. In Figure \ref{fig:DBLPHe-PAu-wd}, paper communities such as ICDM and DAWAK has paired to more than one author communities showing each author community is equally cohesive with these conferences. It can be seen that, distinguished author like George Karypis is associated with ICDM.(some fact about George Karypis) Unlike Maximum weighted bipartite matching, WBC algorithm can give out more than one matching if both matchings have equal weights. This can be seen as authors like Rajeev Rastogi, Minos N. Garofalakis in author community(AU\_10513) are associated with both SIGMOD and VLDB conferences. Similarly 3 author communities(AU\_2241, AU\_9906, AU\_12263) are associated with DAWAK.
On applying MWBC on the CBG created with all Paper and Author communities, we obtained 7 total elements that correspond to the \textit{most cohesive co-authors who also publish frequently in each conference} (shown in Figure \ref{fig:DBLPHe-PAu-wd} with list of few prominent authors.) ICDM and DaWaK have \textbf{multiple author communities} that are \textbf{equally important}. Researchers \textbf{George Karypis} and Michihiro Kuramochi \textbf{are members of one of the frequently publishing co-author groups (in the last 18 years) for ICDM (4 papers)}. Significance of this result is validated from the fact that George Karypis has been a recipient of \textbf{IEEE ICDM 10-Year Highest-Impact Paper Award (2010) and IEEE ICDM Research Contributions Award (2017)}. Moreover, \textbf{multiple conferences can have same cohesive co-author groups}. For example, \textbf{co-authors Rajeev Rastogi and Minos N. Garofalakis are strongly associated with SIGMOD (7 papers) and VLDB (4 papers) in the past 18 years}\footnote{Weights at the layer level are not considered in this analysis. Hence, for an author (e.g., Jiawei Han) who has authored large \textit{number of papers}, his co-authors are distributed among different co-author communities due to lack of weight and hence does not come out. This clearly demonstrates the need for weighted communities at the layer level to increase analysis space as has been shown with meta edge weights.}.

 \begin{figure}[h]
   \centering
%   \vspace{-10pt}
   \includegraphics[width=0.9\linewidth]{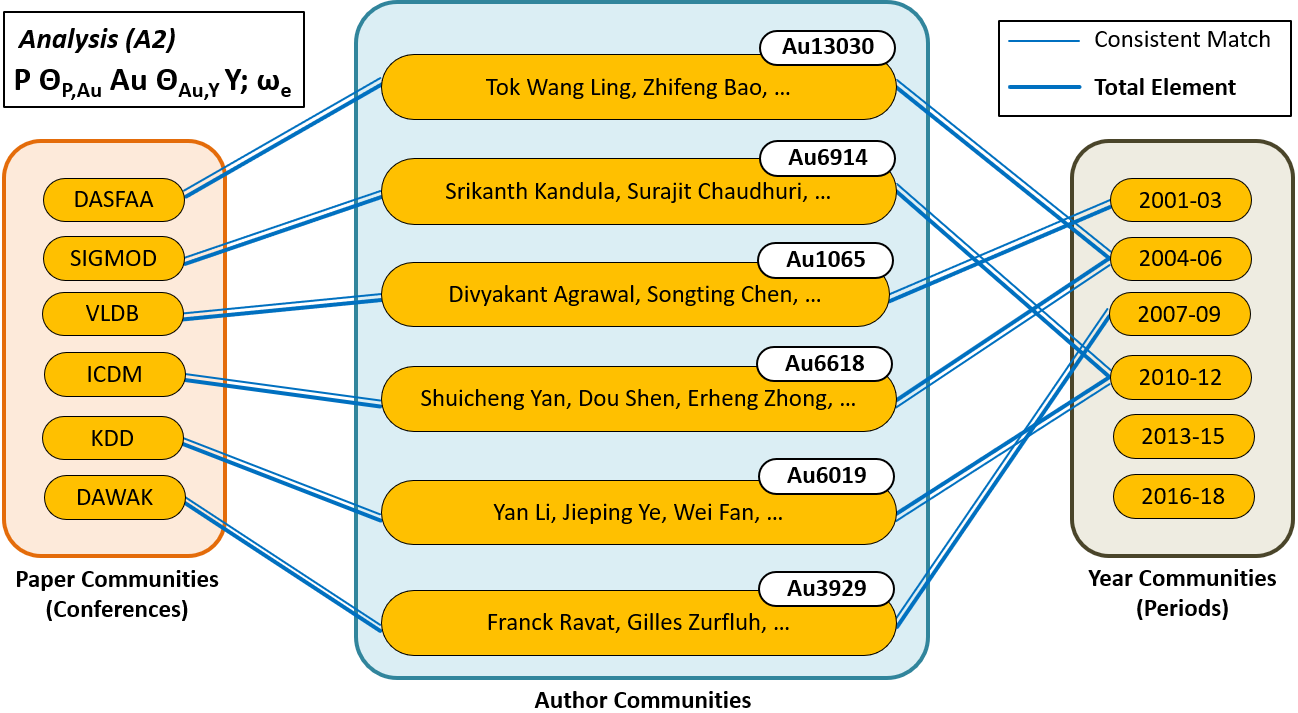}
   \caption{{\ref{analysis:DBLPHe-PAuY} Result}: \textbf{6 Total Elements}}
   \label{fig:A2-DBLP-PAuY-we}
%\vspace{-10pt}
\end{figure}

\noindent\textbf{ \ref{analysis:DBLPHe-PAuY} Analysis: }
%In this analysis, \textit{most popular author communities for each conference was established and for each popular author community we obtained the year in which they were most active in}. In the figure \ref{fig:A2-DBLP-PAuY-we}, each paper communities(conferences) have one most popular group of authors associated. The popular author's group(AU\_13030) of DASFAA and also popular author's group of ICDM(AU\_6618) were most active in the year 2004-2006.
For the required \textit{acyclic 3-community} results, the \textit{most popular author groups} for \textit{each conference} are obtained by MWBC (first composition). The matched 6 author communities are carried forward to find the year periods in which they were \textit{most active} (second composition). 6 total elements are obtained (path shown by \textbf{\textcolor{blue}{bold blue lines}} in Figure \ref{fig:A2-DBLP-PAuY-we}.) %In this analysis, \textit{most popular author communities for each conference was established and for each popular author community we obtained the year in which they were most active in}.
Few prominent names have been shown in the Figure \ref{fig:A2-DBLP-PAuY-we} based on citation count (from Google Scholar profiles.) Clearly, multiple co-author groups can be active in the same year for different conferences as seen from the results. \textbf{For SIGMOD, VLDB and ICDM the most popular researchers include Srikanth Kandula (15188 citations), Divyakant Agrawal (23727 citations) and Shuicheng Yan (52294 citations), respectively who have been active in different periods in the past 18 years}.

An interesting point to be noticed here is that none of the 6 author groups (\textit{obtained from first composition}) had 2013-2015 and 2016-2018 as the most active periods. This is where the relevance of order comes which is derived from the analysis objectives.
%%SC: we have not done this so does not make sense to put it. if we did those ueries, then i would reowrd it differently!
%%We would get some (same or different) matches for the year periods for analysis like \textit{For each co-author community, which is/are the most active periods (Au $\Theta_{Au,Y}$ Y, $\omega_e$)} or \textit{For each period, which is/are most active co-author groups (Y $\Theta_{Y,Au}$ Au, $\omega_e$)}. These results are not shown due to space constraints.

 \begin{figure}[h]
   \centering
%   \vspace{-15pt}
   \includegraphics[width=0.9\linewidth]{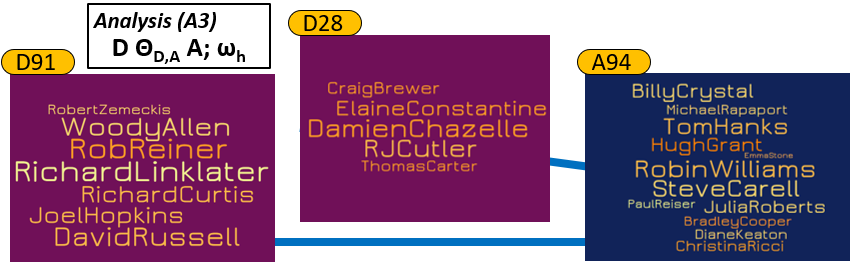}
   \caption{{Sample \ref{analysis:IMdbHe-da-wh} Result} for \textit{Romance, Comedy, Drama}}
   \label{fig:A3-IMDB-DA-wh}
%\vspace{-10pt}
\end{figure}

\noindent\textbf{\ref{analysis:IMdbHe-da-wh} Results: } 34 D-A (Director-Actor) similar genre-based community pairs were obtained, where \textit{majority of most versatile members interact}. Intuitively, a group of directors that prominently makes movies in some genre (say, Drama, Action, Romance, ...) must pair up with the group(s) of actors who primarily act in similar kind of movies. Moreover, a \textit{director group may work with multiple actor groups and vice-versa}. For example, in Figure \ref{fig:A3-IMDB-DA-wh}, the sample result shows that the director groups, D28 and D91, with academy award winners like \textbf{Damien Chazelle and Woody Allen, respectively, pair up with the actor group with members like Diane Keaton, Emma Stone and Hugh Grant}. Members from these groups are primarily known for movies from the \textbf{Romance, Comedy and Drama} genre.

% Due to space constraint, in Fig. \ref{fig:imdb-hemln} (a) we have shown A-D community pairings for the Romance and Comedy genres. Few famous actors and directors from each community have been listed. Such pairings may help production houses to sign up actors and directors for different movie genres. Recently, \textbf{Vin Diesel signed up for Avatar 2 and 3 (Action movie) which is being directed by James Cameroon and this will be the first time they will be collaborating}~\cite{avatar2}. Interestingly, even though they did not work together ever, we paired them together in the groups that corresponded to the Action genre on the basis of \textit{high interaction among other similar actors and directors}. Thus, potential actor-director collaborations can be explored using MLN analysis.

 \begin{figure}[h]
   \centering
%   \vspace{-16pt}
   \includegraphics[width=0.95\linewidth]{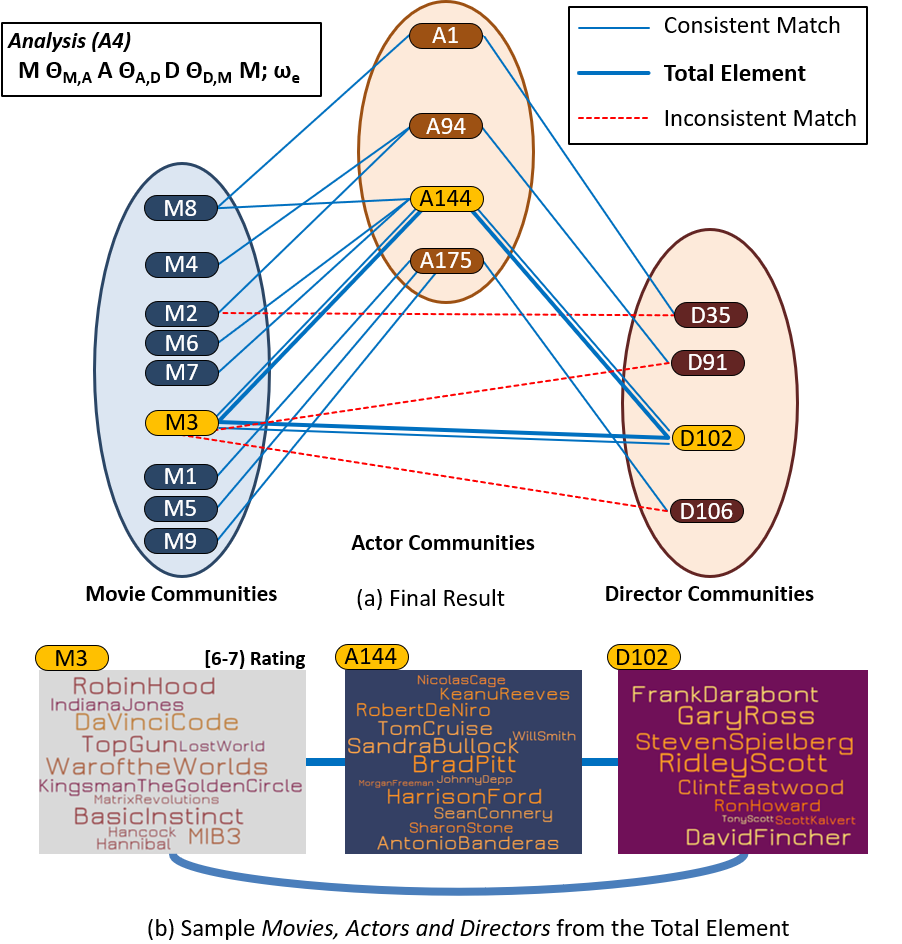}
     %\vspace{-5pt}
   \caption{{\ref{analysis:IMdbHe-madm} Result}: \textbf{1 Total, 9 Partial Elements}}

   \label{fig:A4-IMDB-MADM-we}
%\vspace{-10pt}
\end{figure}
%\kanthi{05/29/2019}{This explanation is as is in VLDB, need to make some changes}

\noindent\textbf{\ref{analysis:IMdbHe-madm} Results: }
Here, the \textit{most popular actor groups for each movie rating class are further coupled with directors}. These \textit{director groups are coupled again with movies to check whether the director groups also have similar ratings}.
Results of each successive pairing (there are 3) are shown in  Figure  \ref{fig:A4-IMDB-MADM-we} (a) using  the  same  color  notation. Coupling of movie and actor communities (first composition) results in 10 consistent matches. %Also, it is easy to see  from  the  figure  that  only  one  of  them  continues  and  becomes  a  total  element  for  the  cyclic 3-community (bold blue triangle.)
When the base case is extended to the director layer (second composition) using all director communities and the matched 4 actor communities, we got 4 consistent matches. The final composition to complete the cycle uses 4 director communities  and  9 movie communities as left and right sets of community bipartite graph,  respectively. \textbf{Only one consistent match is obtained to generate the total element (M3-A144-D102-M3) for the cyclic 3-community (\textcolor{blue}{bold blue triangle}.)} The resulting total element is from the \textbf{Action, Drama genre} as can be seen from the sample members shown in Figure \ref{fig:A4-IMDB-MADM-we} (b). It is interesting to see 3 inconsistent matches (\textcolor{red}{red broken lines}) between the communities  which  clearly indicate  that  all  couplings  are not  satisfied  by  these  pairs. These  result  in  9  partial elements
\textbf{The inconsistent matches also highlight the importance of mapping an analysis objective to a k-community specification for computation.} If a different order had been chosen (viz. director and actor layer as the base case), the result could have included the inconsistent matches. %In this example, we also see \textit{one} no match (\textcolor{blue}{broken blue line}) in the final step, where D12 does not get matched to any movie community, thus generating the partial element, M1-A60-D12.

\subsection{Efficiency of Decoupled Approach}
%plot the number of results for A, M, D, A-M, A-D, ..., A-M-D (all sequences), A-M-D (all sequences with cycle)

\begin{figure}[ht]
   \centering
%   \vspace{-15pt}
  \includegraphics[width=0.9\linewidth]{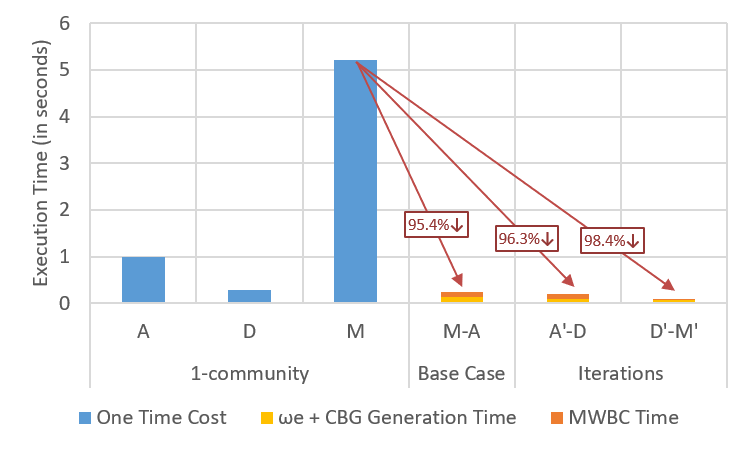}
\vspace{-5pt}
   \caption{Performance Results for cyclic 3-community in \ref{analysis:IMdbHe-madm}}
   \label{fig:perf-madm}
%\vspace{-10pt}
\end{figure}

The goal of the decoupling approach was to preserve the structure as well as improve the efficiency of k-community detection using the divide and conquer approach. We illustrate that with the largest k-community we have computed which uses 3 iterations (including the base case.)
Figure \ref{fig:perf-madm} shows the execution time for the one-time and iterative costs discussed earlier for \ref{analysis:IMdbHe-madm}. The difference in one-time 1-community cost for the 3 layers follow their density shown in Table~\ref{table:IMDbHeMLNStats}. We can also see how the iterative cost is insignificant as compared to the one time cost (by an order of magnitude.) Iteration cost includes creating the bipartite graph, computing $\omega_e$ for meta edges, and MWBC cost. %As the iterations progress, the iterative cost decreases significantly as well.
\textbf{The cost of all iterations together (0.515 sec) is still almost \textit{an order of magnitude less than the largest one-time cost} (5.21 sec for Movie layer.)} We have used this case as this subsumes all other cases. %The zoomed in version (Figure \ref{fig:perf-madm} (b)) of the iterations further show how the iteration cost  reduces. As we had indicated, largest reduction comes for the iteration after the base case which can be clearly seen in Figure~\ref{fig:perf-madm} (b).
%%We believe that the k-community detection can be optimized further.
The \textbf{additional incremental cost for computing a k-community is extremely small validating the efficiency of decoupled approach}.
%%Regarding scalability, each layer can be partitioned to compute 1-community. Also, previously computed results can also be leveraged (e.g., weight computation.)

%%%various components involved in progressively generating the cyclic 3-community for \ref{list:We-cyclic3-madm}. It can be seen from Figure \ref{fig:perf-madm} that the one-time (1-community) cost is bound by the densest layer i.e. movie layer. Moreover, as the number of matches are bound by the base case, thus time taken by the last iteration is the least i.e. Time(Base[M-A]) $<$ Time(Iteration-1[A'-D]) $<$ Time(Iteration-2[D'-M']), where A', D' and M' are the respective sets of communities that are part of some tuple in (k-1)-community. Most importantly, \textbf{the total iteration cost (0.078 seconds) is almost order of a magnitude lower than the one-time cost (0.694 seconds for Movie layer)}. Similar observations were made for all other analysis. Thus,  empirically justifying cost efficiency of the proposed decoupled approach of detecting k-community. 

\section{Conclusions}
\label{sec:conclusions}

In this paper, we have provided a structure- and semantics-preserving definition of a k-community for a HeMLN, its efficient computation, and drill-down of the results.  We proposed a new bipartite-match based composition function that is better-suited for HeMLN Community composition. Finally, we used the proposed approach for demonstrating its analysis versatality using the IMDb and DBLP data sets.

%%we have argued for using multiplexes for modeling as well as analysis. We believe that part of the hesitation to use multiplexes for modeling comes from lack of computation algorithms as compared to other modeling alternatives. Towards that end, we have proposed and developed a community detection approach for HeMLN. We have applied it on the IMDb data set to demonstrate its applicability for flexible analysis as well as computational efficiency using the decoupling approach.

%%Future work includes applying this framework to Homogeneous MLNs. We are also exploring alternate definitions of a MLN community with different analysis characteristics. This decoupling approach also needs to be extended to other analysis concepts, such as centrality detection, subgraph mining, and querying of multiplexes for both types of MLNs.

% or


\ifCLASSOPTIONcaptionsoff
  \newpage
\fi



\bibliographystyle{IEEEtran}
\bibliography{bibliography/santraResearch,bibliography/somu_research,bibliography/itlabPublications}


%\input{texFiles/biography.tex}

\end{document}